\documentclass[11pt]{iopart}

\usepackage{apjfonts}
\usepackage{epsfig}
\usepackage{graphicx}
\usepackage{pifont}
\usepackage{amssymb}
\usepackage{myamsmath}
\usepackage{latexsym}
\usepackage{verbatim}
\usepackage{array}
\usepackage{xspace}
\usepackage{colordvi}
\usepackage{stevedef}
\usepackage{wasysym}
\usepackage{marvosym}
\usepackage{setspace}
\usepackage{iopams}

\newlength{\saveparindent}
\newcommand\myCaption[1]{\small\refstepcounter{figure}%
   \figurename\ \thefigure :\ #1}

\newcounter{saveeqn}%
\newcommand{\alpheqn}{\setcounter{saveeqn}{\value{equation}}%
\stepcounter{saveeqn}\setcounter{equation}{0}%
\renewcommand{\theequation}{\mbox{\arabic{saveeqn}\alph{equation}}}}%
\newcommand{\reseteqn}{\setcounter{equation}{\value{saveeqn}}%
\renewcommand{\theequation}{\arabic{equation}}}%

\begin{document}

\pagenumbering{arabic}
\setcounter{page}{1}

\title[Minimal distance transformations]{Minimal distance transformations between links and polymers: Principles and examples}

\author{Ali R. Mohazab$^\dag$ and Steven  S. Plotkin$^\dag$\footnote{e-mail: steve@physics.ubc.ca}
}

\address{$^\dag$ Department of Physics and Astronomy, University of British
Columbia, 6224 Agricultural Road, Vancouver, BC V6T1Z1, Canada
}



\newpage

\begin{abstract}
The calculation of Euclidean distance between points is generalized to
one-dimensional objects such as strings or polymers. Necessary and
sufficient conditions for the minimal transformation between two
polymer configurations are derived. Transformations consist of
piecewise rotations and translations subject to Weierstrass-Erdmann
corner conditions. Numerous examples are given for the special cases
of one and two links. The transition to a large number of links is
investigated, where the distance converges to the polymer length times
the mean root square distance (MRSD) between polymer configurations,
assuming curvature and non-crossing constraints can be
neglected. Applications of this metric to protein folding are
investigated. Potential applications are also discussed for structural
alignment problems such as pharmacophore identification, and inverse
kinematic problems in motor learning and control.  
\end{abstract}


\submitto{\JPCM}

\maketitle
\normalsize


\newpage
\section{Introduction}
\label{sec:intro}
The standard variational definition of distance can be generalized to
higher dimensional objects such as strings or membranes. 
In a previous paper~\cite{PlotkinSS07}, one of us has introduced the
formalism for this calculation. Consider first zero-dimensional objects (points). 
The distance between two points $A$ and $B$ is
defined through a transformation that takes $A$ to $B$, an object of
dimension one higher than the points themselves (here
one-dimensional). The transformation minimizing the arc-length
travelled between $A$ and $B$ gives the scalar distance
$\mathcal{D}^\ast$. 
The differential increment of arc-length may be defined as either 
$\sqrt{1+(dy/dx)^2 + (dz/dx)^2} dx$, or without the assumption that
$y,z$ are functions of $x$, parametrically.
To be specific, introduce a ``time'' parameter $t$ such that
$0\leq t\leq T$, and 
${\bf r} (0) = {\bf r}_{\mbox{\tiny{A}}}$, ${\bf r}
(T) = {\bf r}_{\mbox{\tiny{B}}}$, 
and 
$\bfr (t) = ( x(t), y(t), z(t) )$. 
The distance between 
${\bf r}_{\mbox{\tiny{A}}}$ and ${\bf r}_{\mbox{\tiny{B}}}$ can be found
variationally~\cite{GelfandIM00}:
\alpheqn
\bea
\mathcal{D}^\ast &=& \mathcal{D}\left[{\bf r}^\ast (t)\right] \mbox{where ${\bf r}^\ast (t)$
  satisfies} \\
&& \d \int_0^T \!\!\! dt \: \left(g_{\mu\nu} \dot{x}^{\mu}(t)
\dot{x}^{\nu}(t) \right)^{1/2} = 0 \: . \label{1dg} \\
\mbox{or} && \d \int_0^T \!\!\! dt \: \sqrt{\rdot^2} = 0 \;\;\;\; \mbox{(Euclidean metric)}
\label{1d}
\eea
\reseteqn
Here we have let  $\dot{x} \equiv dx/dt$, and $\rdot\equiv
d\bfr/dt$. The boundary conditions on the extremal path are 
$\bfr^\ast(0) = \rA$ and $\bfr^\ast(T) = \rB$.

Taking the functional derivative in eq.~(\ref{1d}) gives Euler-Lagrange (EL) equations 
for the Lagrangian $\Lag = \sqrt{\rdot^2}$:
\bea
\frac{d}{d t}\left(\frac{\D \Lag}{\D \rdot}\right) &=& 0 \nonumber \\
\mbox{or} \;\;\;\; \dot{\vhat} &=& 0
\label{eompt}
\eea
with $\vhat$ the unit vector in the direction of the velocity. 

Since the derivative of a unit vector is always orthogonal to that
vector, equation~(\ref{eompt}) says that the direction of the velocity
cannot change, and therefore straight line motion results. Applying
the boundary conditions gives $\vhat =
(\rB-\rA)/\left|\rB-\rA\right|$. However, {\it any} function $\bfv (t)
= \left| \vo(t)\right| \vhat$ satisfying the boundary conditions is a
solution, so long as $\int_0^T \!\! dt \: \left| \vo(t)\right| =
\left|\rB-\rA\right|$. The solution is reparameterization-invariant. 
Then the extremal functional $\bfr^\ast(t)$ is given by
\be
\bfr^\ast(t) = \rA + \frac{\rB-\rA}{\left|
\rB-\rA\right|} \int_0^t \!\! dt \: \left| \vo(t)\right|
\label{eq:rpt}
\ee
and the distance by
\be
\Dist^\ast = \int_0^T\!\! dt \: \sqrt{\rdot^{\ast^2}} = \int_0^T \!\!
dt \: \left| \vo(t)\right| = \left|\rB-\rA\right| 
\ee
which represents the diagonal of a hypercube, as expected. 
At this point we could fix the parameterization by choosing $\left|
\vo(t)\right| =  \left|\rB-\rA\right|/T$ (constant speed), for
example. 

The extremal transformation~(\ref{eq:rpt}) is also a minimum. In
section~\ref{sec:minconds} we will give the sufficient conditions for an
extremum to be a (local) minimum, where we will return to this
example.

The above idea can be generalized to space curves, surfaces, or higher
dimensional manifolds~\cite{PlotkinSS07}. The distance is defined through
the transformation between the objects that minimizes the 
cumulative amount of arc-length travelled by all parts of the
manifold. 

\section{Distance for polymers or strings}
\label{sec:polymer}

Describing the transformation $\bfr(s,t)$ between two space curves $\ra(s)$ and
$\rB(s)$ requires two scalar
parameters: $s$ the arc-length {\it along the space curve}, and $t$ the
``time'' as in the above zero-dimensional case measuring progress
during the transformation. The boundary conditions are then
$\bfr(s,0)=\ra(s)$ and $\bfr(s,T)=\rB(s)$. 
The minimal transformation $\bfr^\ast(s,t)$ 
is an object of dimension one higher
than $A$ or $B$, i.e. it yields a distance that is
two-dimensional. 

The distance $\mathcal{D}^\ast = \mathcal{D}[\bfr^\ast(s,t)]$, where
the functional $\mathcal{D}[\bfr]$ is given by
\be
\mathcal{D}[\bfr] = 
\int_{0}^L\!\!\! ds \!\! \int_0^T \!\!\! dt \: \sqrt{\rdot^2} \: .
\label{dsc}
\ee
Here we have used the shorthand $\bfr \equiv \bfr(s,t) =
(x(s,t),y(s,t),z(s,t))$ (a 3-vector), and
$\rdot \equiv \D \bfr/\D t$.

It has been shown previously that the problem of distance does not map
to a simple soap film, nor to the minimal area of a world-sheet (which
corresponds to the action of a classical relativistic
string)~\cite{PlotkinSS07}. 

Formulated as above, the string can contract and expand arbitrarily in
order to minimize the distance travelled. The transforming object is akin to
a rubber band, and all points on $\ra(s)$ will move in straight lines to
their partner points on $\rB(s)$ to minimize the distance. 
It is worth mentioning that protein
chains for example only change their length by about one percent at
biological temperatures. 
To accurately represent the transformation of a non-extensible string,
a Lagrange multiplier $\lambda(s,t)$ must be introduced into the
effective Lagrangian, weighting the constraint:
\be \sqrt{\bfr'^2} =1 \: ,
\label{constraintL}
\ee
where  $\bfr' \equiv \D \bfr/\D s$.

Under this constraint,
points along the string can no longer move independently of each
other, but must always be a fixed (infinitesimal) distance apart.  The
tangent vector $\tangent=\bfr'$ is now a unit vector, and the total
length of the string is $L = \int_{0}^L\!\! ds \: \sqrt{\bfr'^2} =
\int_{0}^L\!\!  ds$. \\

Consider the minimal distance transformation between two
configurations $\rA(s)$ and $\rB(s)$ of an ideal polymer of length $L$. 
Let us derive the EL equations for this case. 
From equations~(\ref{dsc}) and~(\ref{constraintL}), the effective
action is 
\alpheqn
\bea
\Dist &=&  \int_0^L\!\! \int_0^T \!\!\! ds \, dt \: 
\mathcal{L}\left(\rdot,\bfr'\right) 
\label{D-strings}
\\
\mbox{where} \;\;\;\; \Lag &=&
\sqrt{\rdot^2}-\l \left( \sqrt{\bfr'^2} - 1\right)
\label{L-strings}
\eea
\reseteqn
and the Lagrange multiplier $\l\equiv\l(s,t)$ is a function of both $s$ and
$t$. The extrema of the
distance functional $\Dist$ in~(\ref{D-strings}) are found from
$\d \Dist = 0$. 
Taking the functional derivative gives EL equations~\cite{PlotkinSS07}:
\be
\dot{\vhat} = \l \curv + \l' \tangent \: .
\label{de-EL}
\ee
where $\vhat$ is the unit velocity vector, $\tangent$ is the unit
tangent vector, and $\curv$ is the curvature vector. In
eq.~(\ref{de-EL}) we see explicitly that if the non-extensibility
constraint is set to zero ($\l =0$), all points on $\rA(s)$ 
move in straight lines to $\rB(s)$. 

\subsection{Discrete chains}
\label{sec:discretepolymer}

\begin{figure}
\begin{tabular}{cc} 
\includegraphics[height=0.2\linewidth]{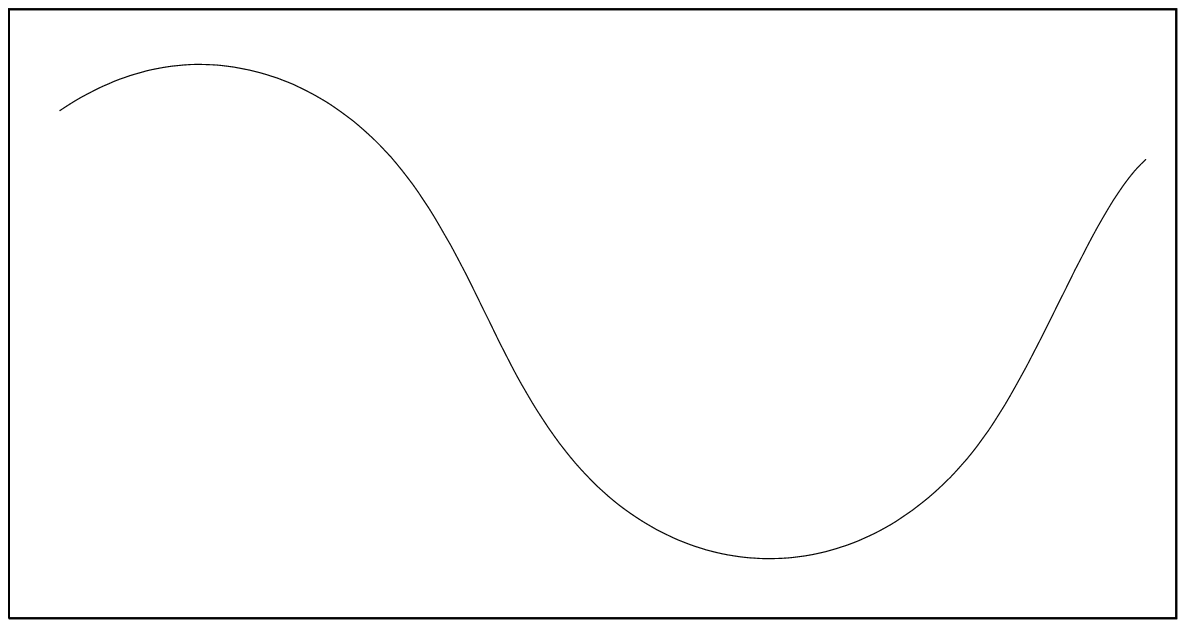} &
\includegraphics[height=0.2\linewidth]{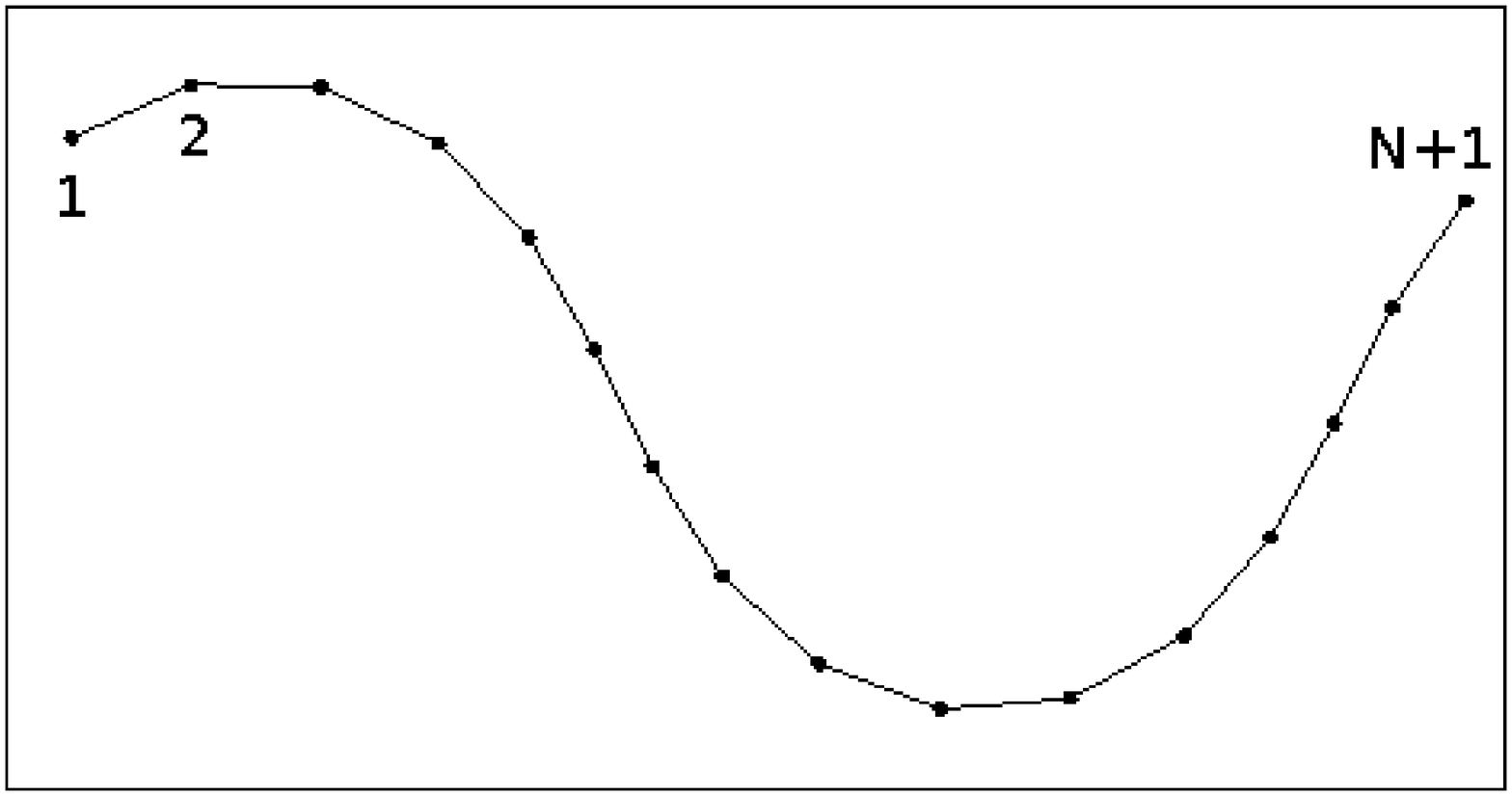}\\
a & b  \\
\end{tabular}
\caption{Continuum (a) and discretized (b) polymer chain. The EL
  equation for the continuum polymer is a nonlinear (vector) PDE, while the EL
  equations for the discretized polymer are a set of nonlinear ODEs.}
\label{fig1}
\end{figure}

To make the problem more amenable to solution, we can discretize the
spatial variables while letting the time variable remain continuous,
i.e. we implement the method of lines to solve eq.~(\ref{de-EL}).
Rather than directly discretizing eq.~(\ref{de-EL}) however, it is
more natural to consider a discretized chain as shown in
figure~\ref{fig1} from the outset, and to calculate the EL equations
for this system.  This recipe then gives the same result as properly
discretizing eq.~(\ref{de-EL}). For the discretized chain, the
constraint in eq.~(\ref{constraintL}) becomes $|\Delta \bfr| = \Delta s
= L/(N-1)$, giving the length of each link.  As the number of beads $N
\rightarrow \infty$ the system approaches a continuous chain. For
finite $N$, the 
Lagrangian becomes a function of the positions and velocities $\left\{
  \bfr_i, \rdot_i \right\}$ of all beads $i$, $1\leq i \leq N+1$. We use
the shorthand notation $\Lag ( \bfr_i, \rdot_i )$.

This recipe yields the distance metric for an ideal, freely-jointed
chain, which has no non-local   
interactions and no curvature constraints. While this approximation
is often used as a first step, real chains may behave
quite differently for several reasons. In many cases, the
configuration which is an energetic minimum is a straight line, or a
single conformation dictated by the chemistry of the polymeric
bonds. At finite temperature, energy in the bath induces
conformational fluctuations. Real polymers also cannot cross
themselves, and because of their stereochemistry also take up
volume. We leave these interesting features for later analysis. 

Equation~(\ref{constraintL}) for the discretized chain becomes $N$
constraint equations added to the effective Lagrangian:  
$$
\sum_{i=1}^{N} \hat{\lambda}_{i,i+1} \left(
  \sqrt{(\bfr_{i+1}-\bfr_{i})^2} - \Delta s \right)
$$
where each $\hat{\lambda}_{i,i+1} \equiv \hat{\lambda}_{i,i+1}(t)$ is a
function of $t$, and $\hat{\l}_{N,N+1} =0$. 
Letting $\l \equiv 2 \hat{\l} \,\Delta s$
and $\bfr_{i+1/i} \equiv \bfr_{i+1} - \bfr_i$
we rewrite this strictly for convenience as 
$$
\sum \frac{\l_{i,i+1}}{2} \: \left( {\bfr_{i+1/i}^2 \over \Delta s^2 }
  -1 \right) \: .
$$

We next convert to dimensionless variables by letting 
$ \bfr = (\Delta s) \hat{\bfr}$. To simplify the notation, 
from here on we simply refer to  $\hat{\bfr}$ as $\bfr$. 
The distance for the discretized chain becomes
\begin{equation}
  \label{eq:discreteD}
  \Dist [ \bfr_i, \rdot_i ] = \Delta s^2 \int_0^T \!\!\! dt \: 
  \Lag\left(\bfr_i , \rdot_i\right) 
\end{equation}
with effective Lagrangian 
\begin{equation}
\label{eq:discreteL}
  \Lag\left(\bfr_i , \rdot_i\right) = 
  \sum_{i=1}^{N} \left( \sqrt{\dot{\bfr}_i^2} - {\l_{i,i+1} \over 2} \left(
      \bfr_{i+1/i}^2 - 1 \right) \right) \: .
\end{equation}

The derivatives $\rdot$ and $\bfr_{i+1/i}$ are raised to different
powers in~(\ref{eq:discreteL}), however so long as $\bfr_{i+1/i}$
satisfies the constraint $\left| \bfr_{i+1/i}\right| = 1$, the EL
equations for $\bfr_i(t)$ will be the same whether the constraint
$\sqrt{\bfr_{i+1/i}^2} = 1$ or $\bfr_{i+1/i}^2 =1$ is used. 

The reparameterization invariance present for point particles
(c.f. section~\ref{sec:intro}) is still present for beads on the
chain, but the parameterization of arclength along the chain is taken
to be fixed by the discretization.

\subsection{General variation of the distance functional}
\label{sec:extremum}

For reasons that will become clear as we progress, we consider the
general variation of the functional $\Dist$, allowing for {\it broken 
extremals}. That is, we allow the curves describing the particle
trajectories to be non-smooth in principle at one or more points in
time. Consider the case of one such point at time $t_1$. The distance
can be written as
\begin{equation}
  \label{eq:Dbroken1}
  \Dist = \int_0^{t_1} \!\!\! dt \: \Lag (\bfr_i,\rdot_i ) + 
\int_{t_1}^T \!\!\! dt \: \Lag (\bfr_i,\rdot_i ) 
\end{equation}

The space-trajectories of the particles must be continuous at time
$t_1$, so $\bfr_i (t_1 - \eps ) = \bfr_i (t_1 +\eps)$, or in shorthand:
\begin{equation}
  \label{eq:rcontin}
  \bfr_i \left(t_1^-\right) =  \bfr_i \left(t_1^+\right) \: .
\end{equation}

\begin{figure}
  \centering
  \includegraphics[height=0.3\linewidth]{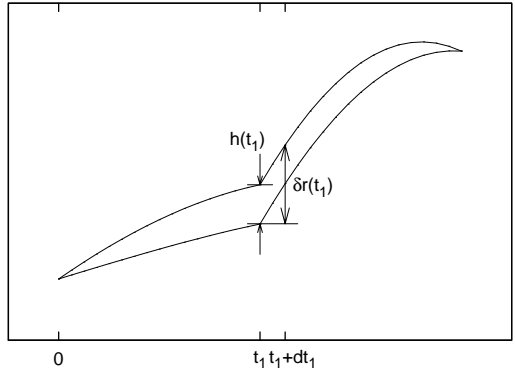}
\caption{General variations of a functional with fixed end points
  allow for broken extremals. In the text we derive the extra
  ``corner'' conditions for a piecewise continuous path to still be
  extremal for our distance functional.}
\label{fig2}
\end{figure}

Let $\bfr_i(t)$ and $\tilde{\bfr}_i (t)$ be two neighboring trajectories
from $\bfr_i(0) = \rAi$ to $\bfr_i(T)= \rBi$ (see figure~\ref{fig2}). 
Neighboring curves will
differ by the first order quantity $\hi(t) = \tilde{\bfr}_i (t)
-\bfr_i(t)$. The fixed boundary conditions at $t=0,T$ dictate that
$\hi(0) = \hi(T)=0$. The difference in distance between the two
trajectories is 
\begin{eqnarray}
  \label{eq:DelD}
  \Delta \Dist &=& \Dist [\bfr_i +\hi ] - \Dist[\bfr_i] \nonumber \\
&=& \int_0^{t_1+\d t_1} \!\!\! dt \: \Lag (\bfr_i + \hi,\rdot_i+\hidot
) - \int_0^{t_1} \!\!\! dt \: \Lag (\bfr_i,\rdot_i )
+ \int_{t_1+\d t_1}^T \!\!\! dt \: \Lag (\bfr_i + \hi,\rdot_i+\hidot)
- \int_{t_1}^T \!\!\! dt \: \Lag (\bfr_i,\rdot_i )
\end{eqnarray}

Taylor expanding the Lagrangian to first order in
$\hi$:${}^{\ddag}$\footnotetext{${}^{\ddag}$ We use the notation
  $F_{\bfr} \equiv \D 
F / \D \bfr$, $F_{\rdot} \equiv \D F / \D \rdot$.}
$$
\Lag \approx \Lag (\bfr_i, \rdot_i ) + \sum_{i=1}^N \left(
  \Lag_{\bfr_i} \cdot \hi + \Lag_{\rdot_i} \cdot \hidot \right)
$$
and integrating by parts using the fixed boundary conditions at
$t=0,T$, the difference in distance up to first order in $\hi$ is
\begin{eqnarray}
  \label{eq:DelDhi}
  \Delta \Dist &\approx& \int_0^{t_1} \!\!\! dt \: \sum_i \left(
    \Lag_{\bfr_i} - {d \over dt} \Lag_{\rdot_i} \right) \cdot \hi
  + \int_{t_1}^T \!\!\! dt \: \sum_i \left( \Lag_{\bfr_i} - {d \over dt}
    \Lag_{\rdot_i} \right) \cdot \hi \nonumber \\
  &+& \Lag (t_1^- ) \d t_1 - \Lag (t_1^+ ) \d t_1 
  + \sum_i \left. \Lag_{\rdot_i} \cdot \hi \right|_{t_1^-}
  - \sum_i \left. \Lag_{\rdot_i} \cdot \hi \right|_{t_1^+}
\end{eqnarray}
with the shorthand $\Lag (t) \equiv \Lag (\bfr_i(t),\rdot_i(t))$. 

\subsection{Conditions for an extremum}
\label{sec:extremumconds}

The variation $\d \Dist$ differs from $\Delta \Dist$ above only by
second order terms. Then for the transformation from $\{\rAi \}$ to
$\{\rBi\}$ to be an extremum, $\d \Dist = 0$.  Thus, the EL equations
(in the top line of eq.~(\ref{eq:DelDhi})) must vanish in each regime
$[0,t_1)$, $(t_1,T]$. Using the form of the Lagrangian in
eq.~(\ref{eq:discreteL}), the EL equations become:
\alpheqn
\begin{eqnarray}
  \label{eq:ELeqnsA}
  &&\vhatdot_1 + \l_{12} \, \bfr_{2/1} = 0 \\
  \label{eq:ELeqnsB}
  &&\vhatdot_2 - \l_{12} \, \bfr_{2/1} + \l_{23} \, \bfr_{3/2} = 0  \\
  \label{eq:ELeqnsC}
  && \hspace{1.0cm} \vdots \nonumber \\
  &&\vhatdot_N - \l_{N-1,N} \, \bfr_{N/(N-1)} = 0 
\end{eqnarray}
\reseteqn

According to equation~(\ref{eq:DelDhi}) there are additional conditions for the
transformation to be an extremum. To find these first note that up to
first order (see figure~\ref{fig2})
\begin{eqnarray}
  \label{eq:hieqn}
  \hi(t_1) \approx \d \bfr_i(t_1) - \rdot_i(t_1) \, \d t_1 \: .
\end{eqnarray}
Then the first variation in the distance is 
\begin{eqnarray}
  \label{eq:cornerD}
  \d \Dist &=& \left[ \left.\left( \Lag - \sum_i \rdot_i \cdot \Lag_{\rdot_i}
      \right) \right|_{t_1^-} - 
    \left.\left( \Lag - \sum_i \rdot_i \cdot \Lag_{\rdot_i}
      \right) \right|_{t_1^+} \right] \d t_1 \nonumber  \\
  &+& \sum_i \left[ \left. \Lag_{\rdot_i} \right|_{t_1^-} - 
    \left. \Lag_{\rdot_i} \right|_{t_1^+} \right] \cdot \d \bfr_i (t_1) 
\end{eqnarray}
which must vanish at an extremum. Because the variations $\d \bfr_i$
and $\d t_1$ are all independent, the terms in square brackets in
equation~(\ref{eq:cornerD}) must vanish. Writing these
expressions in terms of the conjugate momenta $\bfpi = \Lag_{\rdot_i}$
and Hamiltonian $\Ham = \sum_i  \rdot_i \cdot \bfpi - \Lag$ gives the
conditions:
\alpheqn
\begin{eqnarray}
  \label{eq:cornercond1}
&&  \left. \bfpi \right|_{{}_{t_1^-}} = \left. \bfpi \right|_{{}_{t_1^+}} \\
\label{eq:cornercond2}
&& \left. \Ham \right|_{{}_{t_1^-}} = \left.\Ham \right|_{{}_{t_1^+}}
\end{eqnarray}
\reseteqn
These conditions are called the {\it Weierstrass-Erdmann conditions}
or {\it corner conditions} in the calculus of
variations~\cite{GelfandIM00}.

According to the Lagrangian in equation~(\ref{eq:discreteL}), the 
Hamiltonian is given by
$$
\Ham = 
- \sum_{i=1}^{N}  {\l_{i,i+1} \over 2} \left(
      \bfr_{i+1/i}^2 - 1 \right)  
$$
which is identically zero, so corner condition~(\ref{eq:cornercond2})
provides no further information. 

The conjugate momenta according to~(\ref{eq:discreteL}) are given by
\begin{equation}
  \label{eq:conjmom}
  \bfpi = {\rdot_i \over \left| \rdot_i\right| } = \vhat_i \: .
\end{equation}
Therefore, according to corner condition (\ref{eq:cornercond1}),
extremal trajectories cannot suddenly change direction: each $\bfr_i(t)$
follows a smooth path continuous up to first derivatives in the
spatial coordinates. 

The fact that one corner condition provided no information due to the
vanishing of the Hamiltonian is related to our choice of 
parameterization in formulating the problem.
For example, in the case of the distance of the single point particle
mentioned in the introduction, the Lagrangian
may be defined either through independent variable $x$ as 
$\Lag^{(x)} = \sqrt{1+y'^2 + z'^2}$ (with e.g. $y' = dy/dx$), or
parametrically through independent variable $t$ as $\Lag^{(t)}
=\sqrt{\rdot^2}$. The conjugate momenta are then either 
$\Lag^{(x)}_{y'} = y'/\sqrt{1+y'^2 + z'^2}$ and 
$\Lag^{(x)}_{z'} = z'/\sqrt{1+y'^2 + z'^2}$, or
$\Lag^{(t)}_{\rdot} = \rdot/|\rdot| \equiv \vhat$.
The Hamiltonia are either $\Ham^{(x)} = 1/\sqrt{1+y'^2 + z'^2}$ or 
$\Ham^{(t)} = \Lag^{(t)} - \rdot \cdot (\rdot/|\rdot|) = 0$.
The corner conditions can be shown to be equivalent for both choices
of independent variable: for $\Lag^{(t)}$
they give $\vhat (t_1^-) = \vhat (t_1^+)$, so that the direction of
the tangent to the curve cannot have a discontinuity. Together, the
Hamiltonian and two conjugate momenta for $\Lag^{(x)}$ can be
interpreted as components of the unit tangent vector to the curve,
i.e. $\tangent(x) = (\ihat+y'\jhat + z'
\khat)/\sqrt{1+y'^2 + z'^2}$, and so once again the corner conditions enforce a continuous
tangent vector, here $\tangent (x_1^-) = \tangent (x_1^+)$.

\subsubsection{Boundary conditions}
\label{secBCs}

In the continuum limit, the boundary conditions on $\bfr(s,t)$ are
$\bfr(s,0)=\rA(s)$, $\bfr(s,T)=\rB(s)$ where $\rA$ and $\rB$ are the
two configurations of the polymer. For discrete chains, these boundary
conditions become
\alpheqn
\begin{eqnarray}
  \label{eqBCt}
  \left\{\bfr_i(0) \right\} &=& \{\bfr_i^{\left(A\right)} \}  \\
  \left\{\bfr_i(T) \right\} &=& \{\bfr_i^{\left(B\right)} \}  \: .
\end{eqnarray}
\reseteqn

There are also boundary conditions that hold for the {\it end points}
of the chain at all times.  From equations 
(\ref{eq:ELeqnsA}, \ref{eq:ELeqnsC}) we see that there are three
solutions for the end points of the chain:
 
{\bf 1)} If $\l \neq 0$, purely rotational motion results. This can be
seen by taking the dot product of eq.~(\ref{eq:ELeqnsA}) with
$\bfv_1$, which yields $\l_{12} \bfv_1\cdot \bfr_{2/1} = 0$, so the
velocity of the end point is orthogonal to the link. The
rotation must be about a point that is internal to the link, i.e. on
the line between points $1$ and $2$ for end point $1$. This can be
seen straightforwardly for the case of one link by removing point $3$
from equations (\ref{eq:ELeqnsA}) and (\ref{eq:ELeqnsB}). Then the
accelerations $\vhatdot_i$ must be in opposite directions. This can
only occur if rotation is about a point on the line between points $1$
and $2$. 

{\bf 2)} If $\l = 0$, $\vhatdot_i = 0$, and straight-line motion of
the end point results. 

{\bf 3)} Writing out the time-derivative in~(\ref{eq:ELeqnsA}) yields 
\begin{equation}
  \label{eq:vzero}
  \bfv_1^2 \vdot_1 - \left(\bfv_1 \cdot \vdot_1 \right) \bfv_1 =
  -\l_{12} \left| \bfv_1\right|^3 \bfr_{2/1}
\end{equation}
which has the trivial solution $\bfv_1 = 0$. The end point can be at
rest, while other parts of the chain move. 

\subsection{Sufficient conditions for a minimum}
\label{sec:minconds}

For a transformation to be minimal, it is necessary, but not sufficient,
that it be an extremum. We now derive the sufficient conditions
for a given transformation to minimize the
functional~(\ref{eq:discreteD}). 
We describe the formalism in some detail because it is not typically
taught to physicists- for further reading see for example
reference~\cite{GelfandIM00}. 
This section can be read
independently of the others, and might be skipped on first reading. 

According to {\it Sylvester's criterion}, a
quadratic form 
$ \sum_{ij} A_{ij} x_i x_j$
is positive definite if and only if all descending principle minors of
the matrix $\| A_{ij} \|$ are positive, i.e.
\begin{equation}
  \label{eq:sylvester}
  A_{11} > 0 \: , \hspace{0.5cm} 
\begin{vmatrix} A_{11} & A_{12} \\ A_{21} &  A_{22}\end{vmatrix} >
0\: ,  \quad 
\begin{vmatrix} A_{11} & A_{12} & A_{13} \\ A_{21} &  A_{22} & A_{23}
  \\ A_{31} & A_{32} & A_{33} \end{vmatrix} > 0 \: ,
\quad \ldots \quad ,
\mbox{det} \| A_{ij} \| > 0 \: ,
\end{equation}
and a function $F$ of $\bfx \equiv (x_1, x_2, \ldots, x_n)$ has a minimum
at $\bfx^\star$ if the Jacobian matrix $\| \D^2 F/\D x_i \D x_j \|$ is
positive definite at the position of the extremum (where $\D F/\D x_i
= 0$). 

For a function to be a minimum of a given functional, it must satisfy
similar sufficient conditions. Consider again the difference in
distance between two trajectories
in~(\ref{eq:discreteD})${}^{\ddag}$\footnotetext{${}^\ddag$ We
  ignore corner conditions for purposes of the derivation. It can be
  shown that they do not modify the result.}. 
Taylor expanding the Lagrangian to second order in
$\hi$:
\begin{eqnarray}
  \Delta \Dist &=& \Dist \left[ \bfr_i + \hi \right] - \Dist
  \left[ \bfr_i\right] \nonumber \\
 &=& \int_0^{T} \!\!\! dt \: \Lag (\bfr_i + \hi,\rdot_i+\hidot
) - \int_0^{T} \!\!\! dt \: \Lag (\bfr_i,\rdot_i ) \nonumber \\
 &\approx& 
\int_0^{T} \!\!\! dt \: \left[ \sum_{i=1}^N \left(
  \Lag_{\bfr_i} \cdot \hi + \Lag_{\rdot_i} \cdot \hidot \right)
+ {1 \over 2} \sum_{i,j}^{3 N} \left( \Lag_{x_i x_j} h_i h_j + 
2 \Lag_{x_i \xjdot} h_i \hdotj + \Lag_{\xidot \xjdot} \hdoti \hdotj
\right) \right]
\label{eqDmany}
\end{eqnarray}
At an extremum, the first order term in~(\ref{eqDmany}) is zero, and 
$\Delta \Dist \approx \d^2 \Dist$, the
second variation. 
For the extremum to be a minimum, $\d^2 \Dist >0$. 
From eq.~(\ref{eq:discreteL}), the matrix $\| \Lag_{x_i \xjdot}\| =
\|0\|$. Assuming $\| \Lag_{x_i \xjdot} \|$ is in general a symmetric
matrix, i.e. 
$\Lag_{x_i \xjdot} = \Lag_{x_j \xidot}$, the second term in the
quadratic form of~(\ref{eqDmany}) may be integrated by parts to give:
\begin{equation}
  \label{eqDPQ}
  \d^2 \Dist = {1 \over 2} \int_0^T\!\!\! dt \: \left[ 
\braket{\hdot}{\PP \hdot} + \braket{h}{\QQ h} \right] \: ,
\end{equation}
where we have let $\ket{h}$ denote the vector $(h_1,h_2, \ldots ,
h_{3N})$, and 
used the shorthand $\PP$ and $\QQ$ for the matrices:
\begin{eqnarray}
  \label{eqPQ}
  &&\PP(t) = \| \PP_{ij} \| = \| \Lag_{\xidot \xjdot} \|
  \nonumber \\
&&\QQ(t) = \| \QQ_{ij} \| =  \left( \| \Lag_{x_i x_j} \| - 
{d \over dx} \| \Lag_{x_i \xjdot} \| \right) \: .
\end{eqnarray}
From~(\ref{eq:discreteL}) the explicit form for these matrices may be
calculated. $\PP$ is block diagonal:
\begin{equation}
\label{eqPexplicit}
\PP  = 
\begin{bmatrix}
I_{ij}^{\left(1\right)} & 0& \cdots & 0 \\
0 & I_{ij}^{\left(2\right)} & \cdots & 0 \\
\vdots  & \vdots  & \ddots & \vdots  \\
0 & 0 & \cdots & I_{ij}^{\left(N\right)} \\
\end{bmatrix}
\end{equation}
with each block matrix having elements
\begin{equation}
  \label{eqInertia}
  \| \I_{ij}^{\left(J\right)}\| = {1\over
    {\left|\rdot^{\,J}\right|}^3}  
\left( \d_{ij} {\left.\rdot^{\left(J\right)}\right.}^2 - 
\xdot^{\,\left(J\right)}_i \xdot^{\,\left(J\right)}_j
\right)
= {1\over {\left|\rdot^{\,J}\right|}^3}  \begin{bmatrix}
\ydot^2 +\zdot^2 & -\xdot \ydot & -\xdot \zdot \\
-\xdot \ydot & \xdot^2 + \zdot^2 & -\ydot\zdot \\
-\xdot\zdot & -\ydot\zdot & \xdot^2 +\ydot^2 \\
\end{bmatrix}_{\mbox{\footnotesize{(particle $J$)}}}
\end{equation}
Interestingly the numerator of~(\ref{eqInertia}) has the form of an
inertia tensor for a point particle in velocity-space. 
The matrix $\QQ$ is block tri-diagonal, because the spatial derivatives
in~(\ref{eqPQ}) couple each bead to its two neighbors. Using
indices $I,J$ to enumerate beads and $i,j$ to enumerate $x,y,z$
components for each bead:
\begin{eqnarray}
  \label{eqQexplicit}
  & \| \QQ_{IJ,ij} \|& = \d_{ij} \left[ \l_{J-1,J} \left(\d_{IJ} -
      \d_{I,J-1}\right) + \l_{J,J+1} \left( \d_{IJ} -
      \d_{I,J+1}\right) \right]
\nonumber \\
\mbox{or} \hspace{1cm} & \QQ& = 
\begin{bmatrix}
\l_{12} \mathbf{1}  & -\l_{12}\mathbf{1} & 0 & 0& \cdots&   \\
- \l_{12} \mathbf{1} & \left(\l_{12}+\l_{23}\right) \mathbf{1} &
-\l_{23} \mathbf{1} & 0 & \cdots &  \\
0& - \l_{23} \mathbf{1} & \left(\l_{23}+\l_{34}\right) \mathbf{1} &
-\l_{34} \mathbf{1} &  &  \\ 
\vdots  & \ddots &  \ddots  & \ddots & &   \\
&  &  &  & & \\
&  &  &  &  -\l_{N-1,N} \mathbf{1} & \l_{N-1,N}\mathbf{1}   \\
\end{bmatrix}
\end{eqnarray}

For the transformation ${\rstar(t)}$ to be a minimum of $\Dist[\bfr]$,
the functional~(\ref{eqDPQ}) must be positive definite for all
$\ket{h}$. To derive the conditions for this, we can temporarily
ignore the fact that~(\ref{eqDPQ}) arose from the second variation
of~(\ref{eq:discreteD}), and treat~(\ref{eqDPQ}) as a new functional of the function
$\ket{h(t)} = \ket{h_1(t), \ldots h_{3N}(t)}$. We then ask what
$\ket{h(t)}$ extremizes~(\ref{eqDPQ}). 
If $\d^2\Dist >0$ we expect that the only extremal solution would be the trivial
one: $\ket{h(t)}=\ket{0}$, at least for small variations of the
$h_i(t)$. That is, changing the transformation $\{ \rstar_i(t)\}$ from
that which extremized~(\ref{eq:discreteD}) to a neighboring
transformation $\{ \rstar_i(t)+\hi(t)\}$ would increase the
distance travelled. 

The system of $3N$ EL equations for $\ket{h}$ from~(\ref{eqDPQ}) is 
\begin{equation}
  \label{eqELh}
  -{d \over d t} \ket{\PP \hdot} + \ket{\QQ h} = \ket{0}
\end{equation}
with boundary conditions 
\begin{equation}
  \label{eqBCh}
\ket{h(0)}=\ket{h(T)}=\ket{0}  \: .
\end{equation}
Equation~(\ref{eqELh}) is referred to as the {\it Jacobi equation} in the
calculus of variations. 

First note that if $\ket{h}$ satisfies the system of equations
in~(\ref{eqELh}) as well as the boundary conditions~(\ref{eqBCh}),
then integration by parts gives 
\begin{equation}
  \label{eqELhiszero}
  \d^2 \Dist = \int_0^T\!\!\! dt \: \left(
\braket{\hdot}{\PP \hdot} + \braket{h}{\QQ h} \right) = 
\int_0^T\!\!\! dt \: \braket{h}{-{d \over d t} \left( \PP \hdot
  \right) + \QQ h} = 0 \: .
\end{equation}
This means that for $\d^2 \Dist$ to be $>0$, any nontrivial
$\ket{h(t)}$ which 
satisfies the boundary conditions must not itself be an extremal
solution of the Jacobi equation, otherwise solutions $\ket{\rstar(t)}$
perturbed by any constant times $\ket{h(t)}$ are themselves
extremals. One may think of this by analogy as the necessity for the
absence of any ``Goldstone modes'', where excitations by various $C
\ket{h(t)}$ would lead to a family of curves with zero cost in action,
and thus zero effective restoring force,
between them. 

Alternatively we can ask what equation $\h \equiv \ket{h}$ must satisfy if the EL
equations are satisfied for both $\Lag(\bfr,\rdot)$ and the neighboring
extremal $\Lag(\bfr+\h,\rdot+\bfhdot)$. Taylor expanding
$\Lag(\bfr+\h,\rdot+\bfhdot)$ in 
\[
\Lag_{\bfr}(\bfr+\h,\rdot+\bfhdot) - {d \over d t}
\Lag_{\rdot}(\bfr+\h,\rdot+\bfhdot) = 0
\]
gives
\[
-{d \over d t} \left( \Lag_{\rdot\rdot} \cdot \bfhdot \right) + \left( 
\Lag_{\bfr\bfr} - {d \over d t} \Lag_{\bfr \rdot} \right) \cdot \h = 0
\]
which is exactly Jacobi's equation~(\ref{eqELh}) with definitions~(\ref{eqPQ}). 

From here on, it is much simpler to elucidate the central concepts for
sufficient conditions using the case of a single scalar  function
$h(t)$. The analysis can be generalized to the multi-dimensional case
with a bit more effort, but the conclusions are essentially
the same and so they will simply be stated along with the conclusions for the
'1-D' case. For further details see~\cite{GelfandIM00}. 

We write equation~(\ref{eqDPQ}) in 1-D as:
\begin{equation}
  \label{eqPQ1d}
  {1 \over 2} \int_0^T\!\!\! dt \: \left( P \hdot^2 + Q h^2 \right) 
\end{equation}
It was realized originally by Legendre that the integral could be
brought to simpler form by adding zero to it in the form of a total
derivative. Since 
\[
\int_0^T\!\!\! dt \: {d \over dt}\left(w(t) h^2\right) = 0
\]
for any $w(t)$ so long as $h(t)$ satisfies the boundary
conditions~(\ref{eqBCh}), we can add it to the integral
in~(\ref{eqPQ1d}) and seek a function $w(t)$ such that the expression
\[
\d^2 \Dist = {1\over 2} \int_0^T\!\!\! dt \: \left( P\hdot^2 + 2  w h
  \hdot + \left(Q+\wdot\right) h^2 \right)
\]
may be written as a perfect square. This yields the differential
equation 
\begin{equation}
  \label{eqw}
  P\left(Q+\wdot\right) = w^2
\end{equation}
for $w(t)$, and second variation
\begin{equation}
  \label{eqDhsquare}
  \d^2\Dist[h] = {1\over 2} \int_0^T\!\!\! dt \: P \left( \hdot +
    {w\over P} h\right)^2 \: .
\end{equation}

Therefore a necessary condition for a minimum is for $P>0$. The
analogous condition in the multi-dimensional case is for the matrix
$\|\PP\|$ to be positive definite. 

If the differential term $\hdot + {w\over P} h$ in~(\ref{eqDhsquare})
were equal to zero for some $h(t)$, the boundary condition $h(0)=0$
would then imply $\hdot(0)=0$ and thus $h(t)=0$ for all $t$ by the
uniqueness theorem as applied to this first order differential
equation. 

Therefore the functional~(\ref{eqDhsquare}) is positive definite if,
and only if,  \\
{\bf 1.)} $P>0$ ,\\
{\bf 2.)} A solution for eq.~(\ref{eqw}) exists for the whole interval
$[0,T]$. \\
In general, there is no guarantee of condition {\bf (2)} even if
condition {\bf (1)} is valid. For example if $P=1$, $Q=-1$,
(\ref{eqw}) has solution $w(t)=\tan(t+c)$, which has no finite solution if
$|T|>\pi$. $^{\dagger}$\footnotetext{${}^\dagger$ Because reparameterization
  invariance in our problem, the value of $T$ is adjustable, however
  precisely because of this invariance, $\det \|\PP\|=0$ and
  so is no longer positive definite. We discuss this problem and its
  resolution below. }

If~(\ref{eqw}) has a pole at say $\tilde{t}$, then for the
integral~(\ref{eqDhsquare}) to remain finite, $h(\tilde{t})\rightarrow
0$. This point is said to be conjugate to the point $t_o=0$, i.e. it
is a {\it conjugate point}.

Moreover, equation~(\ref{eqw}) is a Riccati equation, which may be brought to
linear form by the transformation $w(t)=-P \Hdot/H$, with $H(t)$ an
unknown function. Substitution in~(\ref{eqw}) gives 
\begin{equation}
  \label{eqELH}
-{d\over dt} \left( P \Hdot\right) + Q H = 0  
\end{equation}
which is precisely equation~(\ref{eqELh})- the Jacobi equation for
$h(t)$. 

This means that for equation~(\ref{eqw}) to have a solution on
$[0,T]$, $H(t)$, as given by the solution to~(\ref{eqELH}), must have
no roots on $[0,T]$. But because equation~(\ref{eqELH}) holds for
$h(t)$ as well, $h(t)$ must have no roots (conjugate points) on
$[0,T]$. Because $h(0)=h(T)=0$, the only way to
extremize~(\ref{eqPQ1d}) is to satisfy eq.~(\ref{eqELH}) with the
trivial solution $h(t)=0$. If $h(t) \neq 0$ for $0<t<T$ then it would
mean that there was a conjugate point at $\tilde{t} =T$. 

In the multi-dimensional case an extremal $\ket{h}$ is one of $3N$
vectors satisfying equations~(\ref{eqELh}), i.e. 
$\ket{h^{(\alpha)}} = \ket{h_1^{(\alpha)}\ldots h_{3N}^{(\alpha)}}$,
$1\leq \alpha \leq 3N$. A conjugate point is defined as a point where
the determinant vanishes:
\[
\det 
\begin{vmatrix}
h_1^{\left(1\right)}\left(t\right) & \cdots & h_{1}^{\left(3N\right)} \left(t\right) \\
\vdots& & \vdots \\
h_{1}^{\left(3N\right)}\left(t\right) & \cdots & h_{3N}^{\left(3N\right)} \left(t\right) 
\end{vmatrix} 
=0
\]
The conditions for a transformation to be minimal are then:\\
{\bf 1.)} The transformation $\ket{\rstar(t)} = \{ \rstar_i (t) \}$ is
extremal, \\
{\bf 2.)} Along $\ket{\rstar(t)}$, the matrix $\PP(t) =
\Lag_{\xidot\xjdot}$ is positive definite, and \\
{\bf 3.)} The interval $[0,T]$ contains no conjugate points to
$t=0$. \\
The above ideas can be made clear with a few examples below. 

\subsubsection{Distance between points}
From the effective Lagrangian $\Lag = \sqrt{\rdot^2}$, 
$\PP = \| \Lag_{\xidot\xjdot} \|$ is given in
equation~(\ref{eqInertia}), which has determinant $\det \PP = 0$, and
so is not positive definite. 
This is due to our choice of parameterization. If we break symmetry by
choosing one spatial direction as the independent variable,
$\Lag\left( x,y',z'\right) = \sqrt{1+y'^2 +z'^2}$ (with e.g. $y'\equiv
dy/dx$ and $x_0 \leq x \leq x_1$). Then 
\[
\PP = {1\over {\left(1 + y'^2 +z'^2\right)^{3/2}}}
\begin{pmatrix} 
1+z'^2 & -y' z' \\  
-y'  z' & 1+y'^2 
\end{pmatrix}
\]
with positive definite determinant $\det \| \PP\| = \left(1+y'^2
  +z'^2\right)^{-1/2} > 0$ for any trajectory. From eq~(\ref{eqPQ}),
$\| \QQ(t)\| = \| 0 \|$. 
Along the extremal, where $y(x) = a x +y_0$, $z(x)=b x + z_0$,
equation~(\ref{eqELh}) gives $\PP\cdot \h' = {\bf c}$, with ${\bf c}$ a
constant vector and $\PP$ a positive definite matrix of constant
values with respect to $x$. Solving this first-order equation gives 
straight line solutions for $\h(x)$. Because $\h(x_0) =0$, there can
be no conjugate points, and because $\h(x_1)=0$, the only solution
to~(\ref{eqELh}) is the trivial one, and the extremum is a minimum. 

\subsubsection{Geodesics on the surface of a sphere}
\label{sec:geodics}

Taking the azimuthal angle
$\phi$ as the independent variable, and polar angle $\theta(\phi)$ as the
dependent variable, the arc-length on the surface of a unit sphere
may be written as   
\be
\Dist[\theta] = \int_{\phi_0}^{\phi_1} \!\!\! d\phi \: \sqrt{\th'^2 + \sin^2\th } \: .
\label{eqDgreatcirc}
\ee
The EL equations give the extremal
trajectory as $\cos\th=A \sin\th\cos\phi + B\sin\th \sin\phi$ with
$A,B$ constants. This is the equation of a plane $z= A x + B y$, which
intersects the surface of the sphere to make a great circle. 
The scalar $P = \Lag_{\th' \th'} = \sin^2\th/\left( \th'^2
  +\sin^2\th\right)^{3/2}$ which is always positive. To
simplify the problem, let $\phi_0=0$, and $\th(\phi_0) = \th(\phi_1)
=\pi/2$, so the great circle lies in the $z=0$ plane. Along this
extremal $P$ is constant and equal to $1$, while $Q=-1$. The second
variation, eq.~(\ref{eqPQ1d}), is then $(1/2) \int_{0}^{\phi_1} \!\!\! d\phi
\: \left( h'^2 - h^2 \right)$. The corresponding Jacobi equation,
$h'' + h =0$, must not have a root between $[0,\phi_1]$. The
nontrivial solution
to the Jacobi equation satisfying the initial condition $h(0)=0$ is
$h(\phi)=C \sin\phi$, which has a conjugate point at $\phi=\pi$. Thus
for the extremal curve to be minimal, 
$\phi_1$ must be  $< \pi$, the location of the opposite pole on the
sphere. If $\phi_1<\pi$, there is no extremal solution for $h(\phi)$
other than the 
trivial one which satisfies the boundary conditions. It is instructive
to look at the arc-length under sinusoidal variations around the
extremal path which satisfy the boundary conditions
$h(0)=h(\phi_1) = 0$, so that $\theta(\phi) = \pi/2 +
h(\phi) = \pi/2 + \eps \sin \left( \pi \phi/\phi_1\right)$. 
Inserting this into eq~(\ref{eqDgreatcirc}) above and expanding to
second order in $\eps$, we 
see that first order terms in $\eps$ vanish, and the difference in
distance from 
the extremal path is $\Delta \Dist =\left( \eps^2 / 4\phi_1 \right)
\left( \pi^2 - \phi_1^2 \right)$. 
For $\phi_1 < \pi$ this is always greater than zero indicating the
extremal is a minimum. For $\phi_1 > \pi$ this is always less than
zero indicating the extremal is a maximum with respect to these
perturbations: the length may be shortened. When $\phi_1=\pi$,
$\Delta \Dist = 0$ to second order. When $h(\phi)$ represents the
difference between great circles $\Delta \Dist$ is precisely zero.

\subsubsection{Harmonic oscillator}

It is not widely appreciated that the classical action for a simple
harmonic oscillator is not always a minimum, and indeed in many cases
can be a maximum with respect to some perturbations. The action for a
harmonic oscillator with given spring constant is proportional to 
$ S[x] = \int_{0}^{T} \!\!\! dt \: {1 \over 2} ( \xdot^2 -
  x^2)$, which has EL equation $\ddot{x} + x =0$. Taking the
specific initial conditions $x(0)=1$, $\xdot(0)=0$, the extremal
solution is $x(t)=\cos t$. The scalar $P(t) = \Lag_{\xdot \xdot} =1$,
which is always positive and satisfies the necessary conditions for a
minimum. The scalar $Q=\Lag_{x x}-{d\over dt} \Lag_{x \xdot} = -1$.
The second variation $\d^2 S[h] = {1\over 2}\int_{0}^{T} \!\!\! dt \:
( \hdot^2 - h^2 )$, which has Jacobi equation
$\ddot{h}+h=0$. This is the same Jacobi equation as that for geodesics
on a sphere, so the sufficient conditions will parallel those above. 
The boundary condition $h(0)=0$ gives $h(t)=A \sin t$,
with conjugate points at $t=n\pi$, $n=1,2,\ldots$. This means that the
action is a minimum only so long as $T<\pi$, i.e. a
half-period. If we let $x(t)$ be the extremal solution plus a $\sin$
perturbation satisfying the Jacobi equation at the conjugate points:
$x(t) = \cos t + \eps \sin t$, then the difference in action from the
extremal path becomes
$ \Delta S = (\eps^2/4 T) (\pi^2 - T^2)$. 
This result is exact because the
action for the oscillator is quadratic (as opposed to the action for
geodesics). 
When $T<\pi$, $\Delta S >0$ indicating the
extremal is a minimum. When $T$ is larger than a half-period, $\Delta
S <0$ and the 
extremal trajectory is a maximum (with respect to half-wavelength
sinusoidal perturbations), and when $T=\pi$, the end point is the
conjugate point and $\Delta S = 0$. 

We discuss sufficient conditions further below in the context of
minimal transformations for links. 

\section{Single Links}
\label{sec:1link}

In the limit of one link, equations
(\ref{eq:ELeqnsA}-\ref{eq:ELeqnsC}) reduce to:
\begin{eqnarray}
  \label{eqELAB}
  &&\vhatdot_{\mbox{\tiny{A}}} + \l \, \bfr_{\mbox{\tiny{B/A}}} = 0
    \nonumber \\ 
  &&\vhatdot_{\mbox{\tiny{B}}} - \l \, \bfr_{\mbox{\tiny{B/A}}} = 0
\end{eqnarray}
where we have let $A$ represent point $1$, $B$ point $2$, and $\l
\equiv \l_{12}$. The link has length $1$ in our dimensionless
formulation, so the vector $\bfr_{\mbox{\tiny{B/A}}}$ could also have
been written as a unit vector $\rhat_{\mbox{\tiny{B/A}}}$.

Both points $A$ and $B$ are end points and satisfy the boundary
conditions of section~\ref{secBCs}. This means that points $A$ and $B$
move by either pure rotation, straight-line translation, or remain at
rest. The initial and final 
conditions may be written $\rA(0) = \bfA$, $\rB(0)=\bfB$, 
$\rA(T) = \bfA'$, $\rB(T)=\bfB'$.

The link in our problem has direction, so $A$ must transform to $A'$
and $B$ to $B'$. We will often use arrowheads in figures to denote
this direction. 

\subsection{Straight line transformations}
\label{sec:stline}

\begin{figure}
\centering
\begin{tabular}{ccc} 
\includegraphics[height=0.2\linewidth]{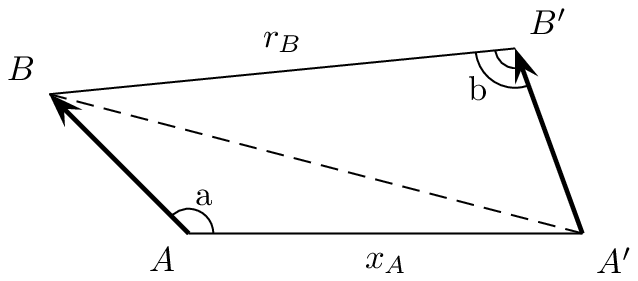}
& \includegraphics[height=0.25\linewidth]{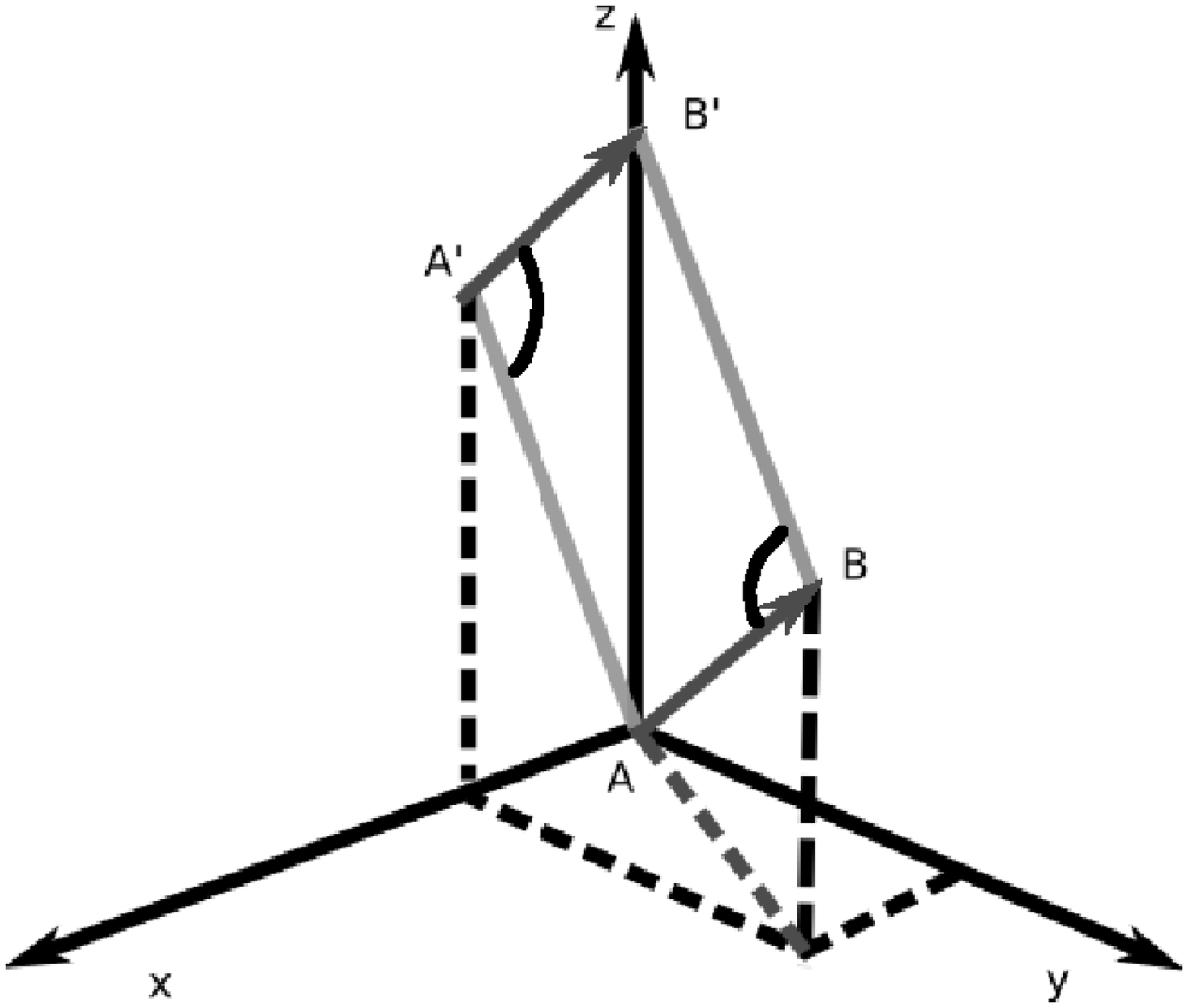}
& \includegraphics[height=0.15\linewidth]{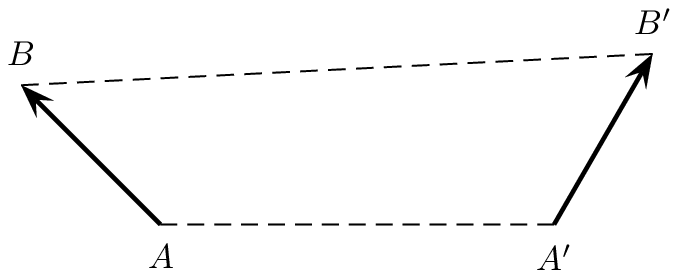}
\\ 
a & b & c \\
\end{tabular}
\caption{Possible (a,b) and impossible (c) straight line transformations
  between links $AB$ and $A'B'$. Figure $b$ shows a straight line
  transformation where the initial and final states do not lie in the
  same plane. In the text we derive the conditions for the possibility
  of a straight line transformation between links. }
\label{obtuse-impossible}
\end{figure}

As a first example, consider the two links shown in
figure~\ref{obtuse-impossible}a. 
The four points $A,B,A',B'$ need not lie in a plane (see for example
fig~\ref{obtuse-impossible}b).
Let angle $\angle BAA' \equiv a$ be obtuse. 
We draw straight lines from $A$ to $A'$ and $B$ to $B'$, and ask
whether such a transformation is possible. 
We can thus derive the following rule: \\
$\bullet$ {\it For a straight line transformation to exist between two
  links, opposite angles of the quadrilateral made by $\overline{AB}$, 
$\overline{A'B'}$, $\overline{AA'}$, $\overline{BB'}$ must be obtuse.}

Let the length that point $A$ travels be $\xA$, i.e. we imagine the
point $A'$ and the distance $\xA=|AA'|$ to be variable. The
length $\rb$ that 
point $B$ travels is then a function of $\xA$ and the original angle
$a$, $\rb(\xA,a)$. We can now find conditions on the angle $b\equiv
\angle BB'A'$ such that the transformation is possible. 

After some distance $\xA$ travelled by point $A$, the length of the
line from $B$ to $A'$ is
\begin{eqnarray}
\overline{BA'} &=& \xA^2 + 1 - 2 \xA \cos a \nonumber \\
&=& \rb^2 + 1 - 2 \rb \cos b \nonumber    
\end{eqnarray}
so that 
\[
\rb(\xA,a) = \cos b \pm \sqrt{ \cos^2 b + f(\xA,a) }
\]
with $f(\xA,a)=\xA^2 -2\xA\cos a$. Since $a$ is obtuse, $f>0$ when
$\xA>0$, and so 
the positive root must be taken for $\rb$ to positive.
When $\xA=0$, $f(0,a)=0$, and 
\[
\rb(0,a)=\cos b + \left| \cos b \right| = 0
\]
Therefore $b$ must also be an obtuse angle. If two opposite angles are
obtuse, then the other two angles must be acute. This concludes the
proof that the above conditions are sufficient. An additional proof
that they are necessary is given in~\ref{app:necessary}.

\begin{figure}
\centering
\begin{tabular}{cc} 
\includegraphics[height=0.2\linewidth]{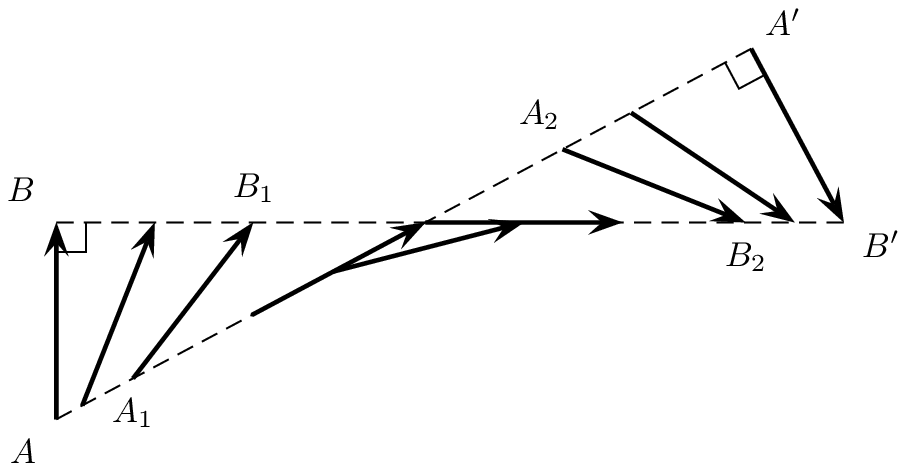} &
\includegraphics[height=0.3\linewidth]{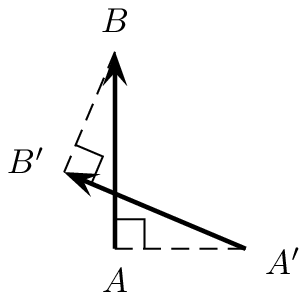} \\
A &  B 
\end{tabular}
\myCaption{\baselineskip 10pt \\
(A) An example of a set of link configurations connected by a
  straight-line transformation. The link rotates clockwise as
  it translates to allow the end points to move in straight lines. The
translation can proceed no farther than the end points $AB$ and
$A'B'$, which have link vectors $\ora{AB}$ or
$\ora{A'B'}$ that are 
perpendicular to one or the other of the vectors $\vhatA$ or
$\vhatB$. The totality of states thus connected forms a ``bowtie''. 
(B) A bowtie where the terminal states $AB$ and $A'B'$ happen to cross
each other.}
\label{bowtie}
\end{figure}

We readily see that figure~\ref{obtuse-impossible}A is one pair of a
larger set of 
straight line transformations that can continue until one or both of
the obtuse angles reaches $90^\circ$. This collection forms a
``bow tie'' of admissible configurations, as in figure~\ref{bowtie}. 
Note that straight lines in the quadrilateral may cross as in the
transformation from $A,B$ to $A', B'$ in figure~\ref{bowtie}. 
Trivial translations of the link without any concurrent rotation are a
special case of general straight line transformations. 

\subsection{Piece-wise extremal transformations: transformations with rotations}
\label{sec:piecewise}

An immediate question is the nature of the transformation between $AB$
and $A'B'$ in figure~\ref{obtuse-impossible}C, where opposite angles of the
quadrilateral are not obtuse. 
Recall our link has direction so $A$ cannot transform to $B'$.
Then direct straight-line solution is not
possible due to the constraint of constant link length.

The only remaining solution is for the link to rotate as part of 
the transformation. 
Consider first the rotation of
link $AB$. 
The EL equations~(\ref{eqELAB}) allow for pure rotations about $A$,
$B$, or a common center along the link. 
Likewise for link $A'B'$. 

\begin{figure}
\centering
\begin{tabular}{|cc|cc|} \hline
a&&b& \\
 & \epsfig{file=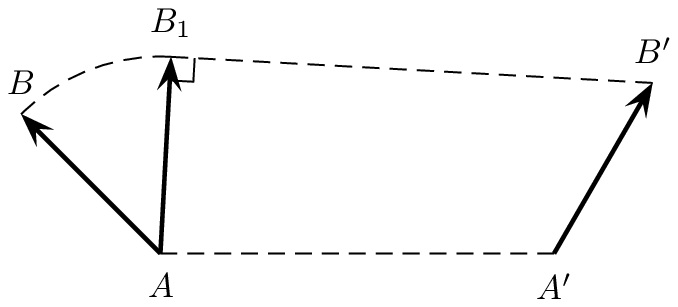,width=0.3\linewidth,clip=} &  &
\epsfig{file=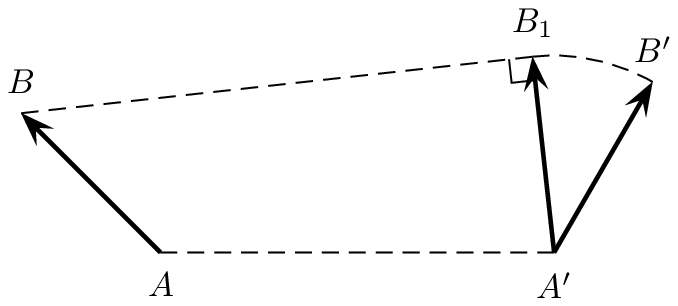,width=0.3\linewidth,clip=}  \\
\hline
c && d & \\
 &  \epsfig{file=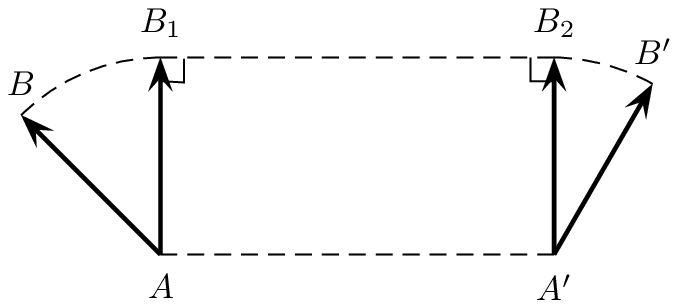,width=0.3\linewidth,clip=}  &  & 
\epsfig{file=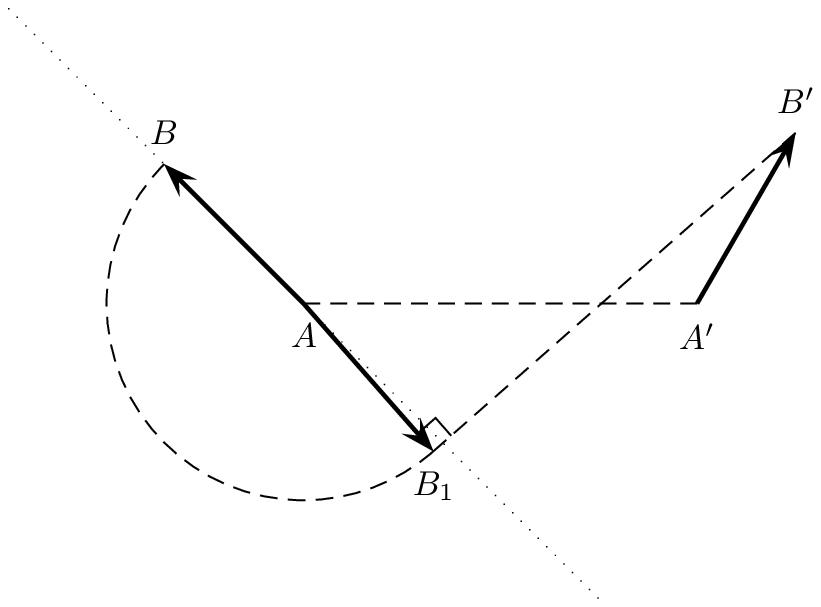,width=0.3\linewidth,clip=}  \\
\hline
e && & \\
& \epsfig{file=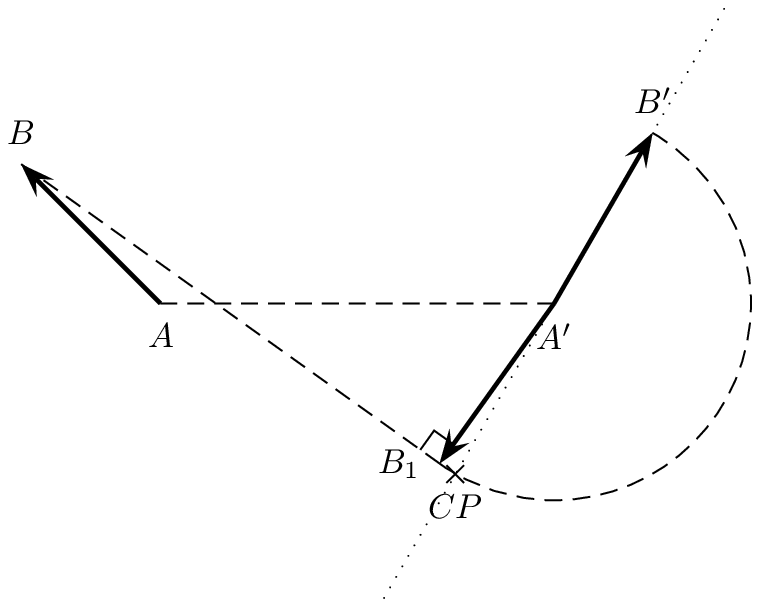,width=0.3\linewidth,clip=} &
& 
\\
\hline
\end{tabular}
\caption{Transformations between two links involving broken extremals
  consisting of 
  rotation and translation. $(b)$ is the global minimum, with shortest
  distance travelled during the transformation. $(a)$, $(c)$, and
  $(d)$ are local 
  minima. $(e)$ is 
  extremal, but not minimal as the trajectory of arc
  $\stackrel{\frown}{B'B_1}$ passes through a conjugate point.}
\label{rotfigs}
\end{figure}

The rotation can occur from either link $AB$ (fig~\ref{rotfigs}a) or
link $A'B'$ (fig~\ref{rotfigs}b).  
After the link rotates to a critical angle, it can then travel in a
straight line. 
The extremals are broken in that they involve matching up a piece
consisting of pure rotation with a piece consisting of pure
translation of the end points of the link. Where the pieces match they
must satisfy the corner conditions~(\ref{eq:cornercond1},
\ref{eq:cornercond2}). This means that the end points cannot suddenly
change direction, a situation which is only satisfied by a straight
line trajectory that lies tangent to the circle of rotation.

From figure~\ref{bowtie}, we see that a straight line
transformation exists only when an angle between a link and one of the
straight line trajectories reaches $\pi/2$. 

The critical
angle that link $AB$ must rotate is then determined by the point where a
line drawn from $B'$ is just tangent to the unit sphere centered at
point $A$, point $B_1$ in figure~\ref{rotfigs}a. There is generally a different
critical angle if the rotation occurs at link $A'B'$ as in
fig~\ref{rotfigs}B. It is shown in~\ref{app:critangles} that
in general  
the critical angle is determined by drawing the tangent to a circle or
sphere about one of the link ends.

If the rotation was about a common center, we see that one or
another of the link ends would violate a corner condition, so the rotation
must be about one of the link ends.

According to eqs.~(\ref{eqPexplicit}) and~(\ref{eqInertia}), the
matrix $\PP$ has a determinant of zero due to the parametric
formulation in the problem and so is not positive definite.
To show that the transformations in fig.~\ref{rotfigs}a,b are indeed
minimal, we need to then express the problem in non-parametric form. 
To do this, let the independent variable be the angle $\th$ of the
link with the vertical. Then the displacement $x$ along the line
$\overline{AA'}$ is the unknown function of $\th$ to be determined by
minimizing the total arc length travelled. This distance can be
written as 
\[
\Dist [x] = \int_{\th_0}^{\th_1} \!\!\! d\th \: \left( \sqrt{x'^2 + 2 x'
  \cos \th + 1} + \sqrt{x'^2} \right)
\]
In this formulation, the scalar quantity $\PP(\th) = \Lag_{x' x'}$
becomes
\[
\PP(\th) = {\sin^2\th \over \left( x'^2 + 2 x' \cos \th +
    1\right)^{3/2} }
\]
which is always $>0$ except for the isolated point $\th=0$, in
particular it is positive along the extremal trajectory which is
necessary for a minimum.
So we conclude that the transformation with the smaller angle of
rotation in fig~\ref{rotfigs}b
is here the global minimum, and the other transformation
(fig~\ref{rotfigs}a) is a local minimum. 

Figure~\ref{rotfigs}e is also an extremal trajectory, satisfying corner
conditions, and with positive definite $P$. However it is not a local
minimum because the trajectory passes through a conjugate
point (denoted by point $CP$, where the dotted line along
$\overline{A'B'}$ meets the great circle about $A'$). 
According to the results in section~\ref{sec:geodics}, if the
extremal trajectory (a great circle) traverses an angle larger than
$\pi$ radians, it 
passes through a conjugate point and thus becomes unstable
to long-wavelength perturbations. Transformations involving rotations
about points $B$ or $B'$ 
in figure~\ref{rotfigs} both have conjugate points and so are not
minimal. 

The transformation in fig.~\ref{rotfigs}c does not pass through a
conjugate point and so is in fact another local minimum.  The part of
the extremum
along the straight line section of the trajectory has no conjugate
points as discussed above.

\subsection{Systematically exploring transformations by varying link positions}

\begin{figure}
\centering
\begin{tabular}{|cc|cc|cc|} 
\hline
a && b & & c & \\
 & \epsfig{file=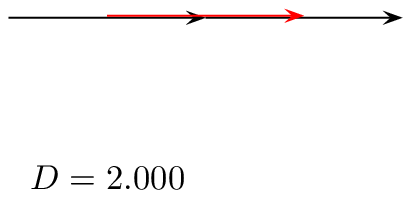,width=0.3\linewidth,clip=} &  &
\epsfig{file=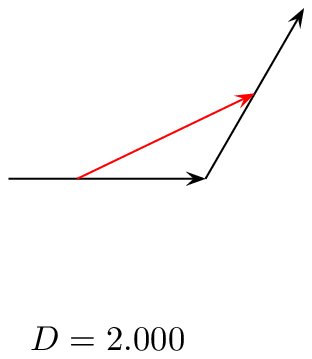,width=0.3\linewidth,clip=}   &  &
\epsfig{file=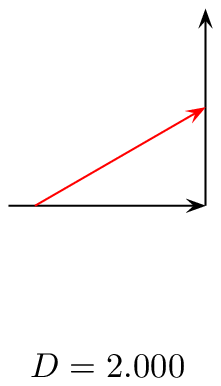,width=0.3\linewidth,clip=}   \\
\hline
d && e & & f & \\
 & \epsfig{file=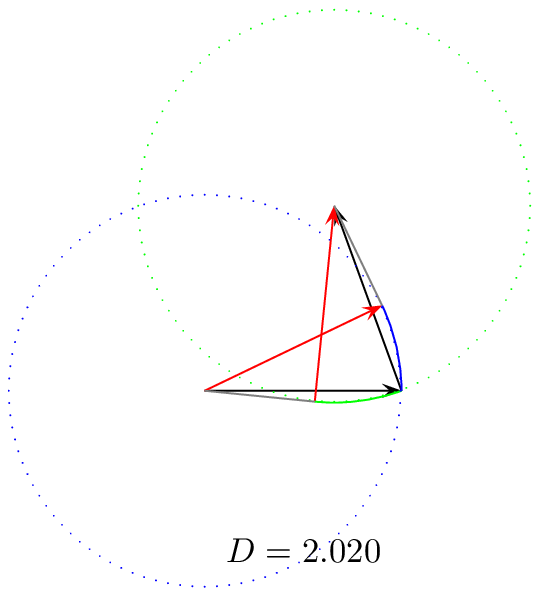,width=0.3\linewidth,clip=} &  &
\epsfig{file=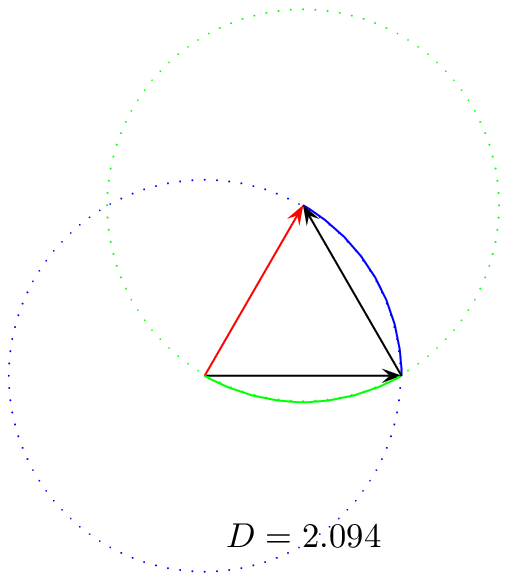,width=0.3\linewidth,clip=}   &  &
\epsfig{file=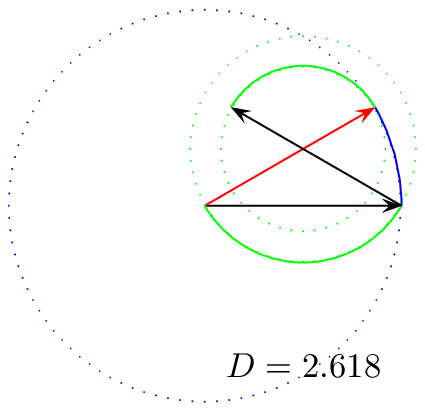,width=0.3\linewidth,clip=} \\
\hline
g && h & &  & \\
 & \epsfig{file=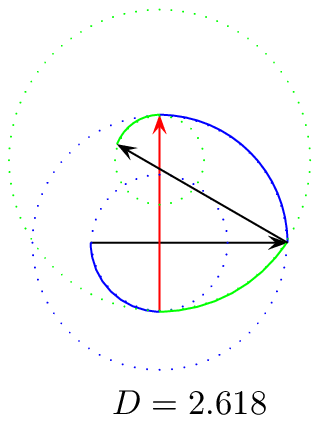,width=0.3\linewidth,clip=} &  &
\epsfig{file=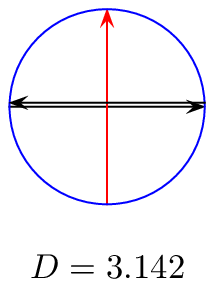,width=0.3\linewidth,clip=}   &&
\\
\hline
\end{tabular}
\caption{Successive transformations between two links made by rotating
  a link so that there is a progressively larger angle between
  the links as vectors (or smaller angle made between them as lines). The
  two boundary conditions (the initial and final conditions) are shown
  as black links, and an  
  intermediate state is shown as a red link or links. The arcs traced
  out by the end points are shown in blue or green, while straight line
  motions when they are not along the links themselves are shown in
  grey. The distance
  travelled over the course of the transformation is given 
  below each figure.} 
\label{systransf}
\end{figure}

We can investigate what happens to the minimal transformation when one
of the link positions or angles is varied with respect to the other. 
Let us start by putting the two links head to tail as shown in
figure~\ref{systransf}A. The distance between them is $2$ by simple
translation of link end points. 

We can now increase the angle between the two vectors by rotating the
right link for example, as in figures~\ref{systransf}B-H. So long as the
angle between the two vectors is less than $90^\circ$, one link may
slide along another and the distance is unchanged
(figs~\ref{systransf}A-C). This is a special case of the transformations
shown in figure~\ref{bowtie} (compare for example
figure~\ref{systransf}B with the 
middle three unlabelled links in that figure). 

Beyond $90^\circ$ however, the transformation must include
rotation. Fig~\ref{systransf}D has an angle of~$150^\circ$. The minimal
transformation first rotates, for example with the tail of the
horizontal black arrow fixed, and the
head tracing out the blue arc, until the critical angle is reached,
where a straight line made from the final arrowhead (at the top of the
figure) is just tangent to the circle made by the blue arc. This state
is indicated by a red link in figure~\ref{systransf}D. The link then 
translates to its reciprocal position at the opposite end of the
bowtie, denoted by a second red link (c.f. also figure~\ref{bowtie}B). 
At this point the arrowhead has completed the transformation. Finally
the tail rotates into its final position.
The total distance travelled is slightly larger than $2$. 

When the angle between the vectors is $120^\circ$ as shown
in~\ref{systransf}E, the transformation consists of pure
rotations. Taking the initial state to be the horizontal black vector,
the link first rotates about its fixed tail, the head tracing out the
blue arc, until the link reaches the state shown in red, where the
position of the arrowhead has reached its final end point. Then the
link rotates about its head until the position of the tail reaches the
final state. 

When the angle between the links is larger than $120^\circ$ as shown
in figs~\ref{systransf}F-G, the transformation must involve rotation
about an internal point along the link. Let points $A$ and $B$ denote
the tail and head of the link respectively. If an infinitesimal
rotation $\Delta \th$ occurs about an internal point $P$, the
increment in distance travelled is
\[
\Delta \Dist = | \bfr_{\mbox{\tiny{B/P}}} | \Delta \th 
+ | \bfr_{\mbox{\tiny{B/A}}} | \Delta \th = \Delta \th
\]
which is {\it independent} of the position of the instantaneous center
of rotation (ICR). This means that there are an infinity of
transformations all giving the same distance, depending on the
time-dependence of the ICR. Two simple alternatives with only two
discrete positions of ICR are shown in
figures~\ref{systransf}F,G. Specifically, in figure~\ref{systransf}F, the
horizontal black vector first rotates about its tail to the red
configuration, which is a mirror image of the final black vector. 
Then rotation is about an internal point determined by
the intercept of the red vector with the final black vector, with end
points tracing out the green arcs. In figure~\ref{systransf}G the two
ICRs are both internal and determined by the intercepts of the initial
and final states with the red vector shown. 

Figure~\ref{systransf}H depicts the transformation for overlapping, opposite
pointing vectors. Rotation can now only occur about one point in the
center of the vectors. 

\begin{figure}
\centering
\begin{tabular}{|cc|cc|} 
\hline
 &a &  &e \\
 & \epsfig{file=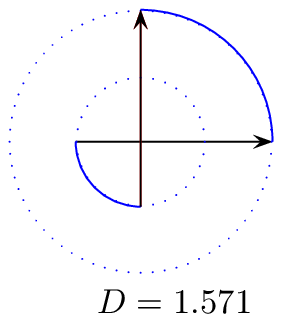,width=0.3\linewidth,clip=} &  &
\epsfig{file=Figs/1link-5.eps,width=0.3\linewidth,clip=}   \\
\hline
 &b &  & f \\
 & \epsfig{file=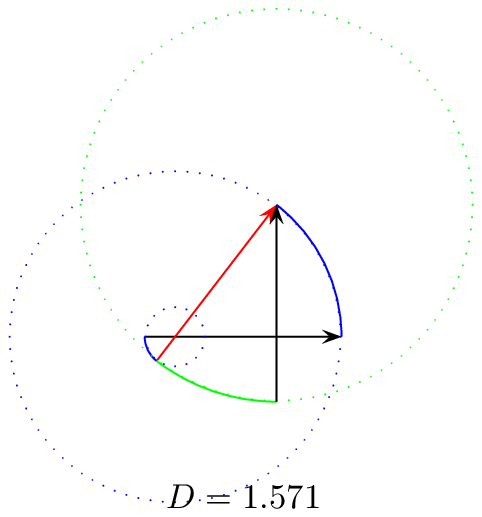,width=0.3\linewidth,clip=} &  &
\epsfig{file=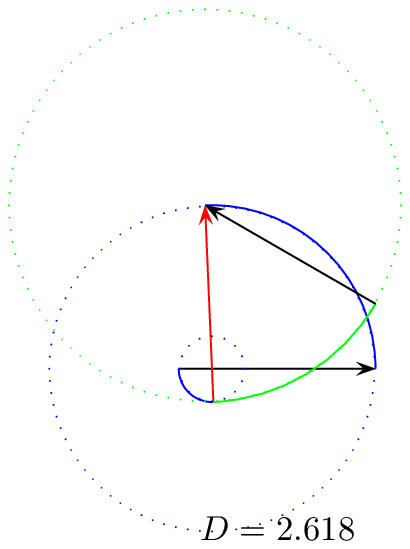,width=0.3\linewidth,clip=} \\
\hline
&c & & g  \\
 & \epsfig{file=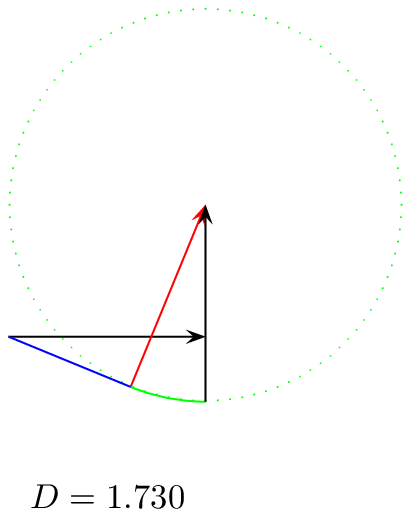,width=0.3\linewidth,clip=} & &
\epsfig{file=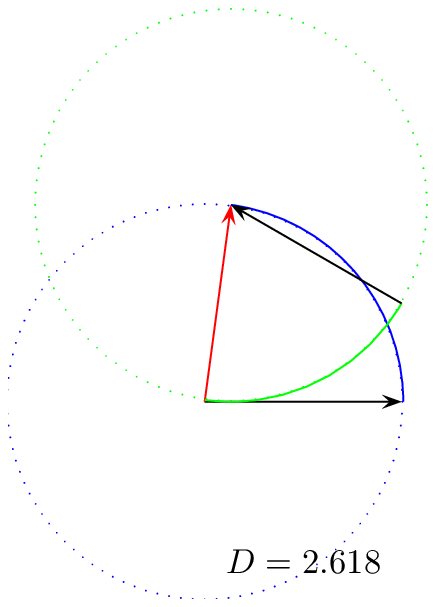,width=0.3\linewidth,clip=} \\
\hline
&d & & h  \\
& \epsfig{file=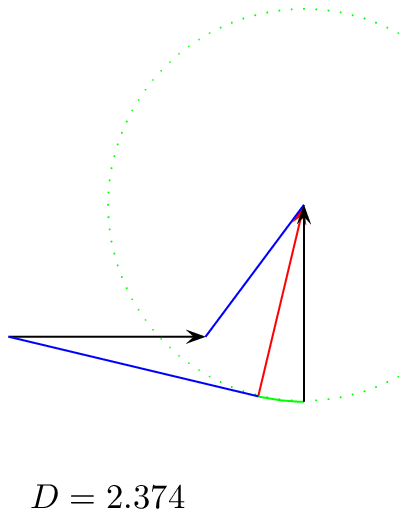,width=0.3\linewidth,clip=} &  &
\epsfig{file=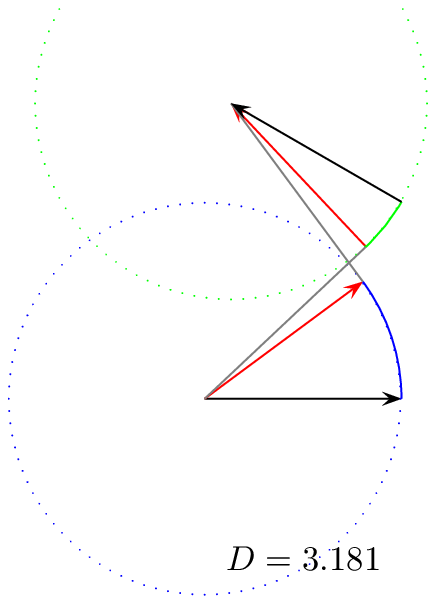,width=0.3\linewidth,clip=}   \\
\hline
\end{tabular}
\caption{Successive transformations between two links made by
  translating one link with respect to the other. In (a-d) the initial
and final configurations are perpendicular, while in (e-h) they are at
an angle of $150^{\circ}$ to each other. Note the distances in (e-g) are all
the same, even though the end points of the links are at varying
distances from each other. }
\label{systrans2}
\end{figure}

Figure~\ref{systrans2} illustrates what happens when one of the links
is translated with respect to another, starting from two different
scenarios shown in~\ref{systrans2}A and~\ref{systrans2}E.
In~\ref{systrans2}A, the tail of the vertical link is displaced
$(1/3,-1/3)$ with respect to the tail of the horizontal link. The
minimal transformation is a pure rotation by $\pi/2$. 

In figure~\ref{systrans2}B, the tail of the vertical link is now
displaced to
$(2/3,-1/3)$. 
Pure rotations again give a distance of $\pi/2$. Rotation about a
point on the horizontal link that is equidistant from both arrowheads
transforms the initial arrowhead to the final (red intermediate
state). Then rotation of the tail about the arrowhead transforms to
the final state. 

In figure~\ref{systrans2}C, the minimal transformation first involves
a translation by sliding the arrowhead along the vertical, until the
arrowheads overlap (red intermediate state). The tail end of the link
then rotates into place. 

In figure~\ref{systrans2}D, straight lines from the end points will
not satisfy the obtuse condition in section~\ref{sec:piecewise}, so
the transformation must involve rotations. Here a straight line
transformation takes the link almost to the final state. It then must
undergo a small rotation to complete the transformation. Seen in
reverse, the vertical arrow must rotate to a critical angle determined
by the criterion in section~\ref{sec:piecewise}, before the link can
finish the transformation by pure translation. 

Figure~\ref{systrans2}E is figure~\ref{systransf}F once again. The
final condition (the tilted link) will be systematically changed by
translating it vertically away from the horizontal link (which we
choose arbitrarily as the initial configuration). 

In figure~\ref{systrans2}F the tilted link is translated a distance
$1/3$ vertically. The transformation can be achieved by rotating the
horizontal link about a point equidistant from both arrowheads, to the
red intermediate configuration. The link then rotates about the
arrowhead into the final configuration. The distance is still the
angle rotated for the reasons mentioned above in the context of
figures~\ref{systransf}F-G, $\th = (150/180)\pi$, which is unchanged
from~\ref{systrans2}E. In fact, so long as the arrowhead can be
reached by rotation (the translated distance is less than $d$ where
$d$ is the solution to $d^2+d+1-\sqrt{3}=0$ for this angle), then the
distance will be unchanged. The transformation at the critical
distance is shown in figure~\ref{systrans2}G. The rotations now occur
about the end-points: the tail and head of the link. 

In figure~\ref{systrans2}H the translated distance is now equal to
$1$. The transformation first consists of a rotation about the tail to
a critical angle (blue arc and red intermediate state), then a
translation much like that in figure~\ref{bowtie} (grey straight lines
between red intermediate states), and finally a rotation about the
head (green arc) to the final configuration.

\section{2-link chains}
\label{sec:2link}

We now consider the next simplest case of $2$ links ($3$ beads). The
Lagrangian now reads: 
\begin{equation}
  \label{eq:L3}
  \Lag (\bfr_1, \bfr_2, \bfr_3, \dot{\bfr}_1, \dot{\bfr}_2,
  \dot{\bfr}_3 ) = \sqrt{\dot{\bfr}_1^2}+
  \sqrt{\dot{\bfr}_2^2}+ \sqrt{\dot{\bfr}_3^2} 
  -{1\over2}\l_{12} \left( \left(\bfr_2 - \bfr_1 \right)^2 -1 \right)
  -{1\over2}\l_{23} \left( \left(\bfr_3 - \bfr_2 \right)^2 -1 \right)
\end{equation}
which has EL equations (c.f. eq.s~\ref{eq:ELeqnsA}-\ref{eq:ELeqnsC})
${}^{\ddag}$\footnotetext{${}^{\ddag}$ The links have length $1$ in our
  dimensionless 
formulation, so the vectors $\bfr_{\mbox{\tiny{B/A}}}$ and
$\bfr_{\mbox{\tiny{C/B}}}$ could also have 
been written as unit vectors $\rhat_{\mbox{\tiny{B/A}}}$ and
$\rhat_{\mbox{\tiny{C/B}}}$.} :
\alpheqn
\begin{eqnarray}
  \label{eq:ELeqnsA2link}
  &&\vhatdot_{\mbox{\tiny{A}}} + \l_{\mbox{\tiny{AB}}}  \,
  \bfr_{\mbox{\tiny{B/A}}} = 0 \\ 
  \label{eq:ELeqnsB2link} 
  &&\vhatdot_{\mbox{\tiny{B}}} - \l_{\mbox{\tiny{AB}}} \,
  \bfr_{\mbox{\tiny{B/A}}} +  \l_{\mbox{\tiny{BC}}} \, 
\bfr_{\mbox{\tiny{C/B}}} = 0 \\
  \label{eq:ELeqnsC2link}
  &&\vhatdot_{\mbox{\tiny{C}}} - \l_{\mbox{\tiny{BC}}}
  \,\bfr_{\mbox{\tiny{C/B}}}   = 0  \: .
\end{eqnarray}
\reseteqn

The corner conditions~(\ref{eq:cornercond1}),~(\ref{eq:conjmom}) imply
\[
\vhat_i \left( t^- \right) = \vhat_i \left( t^+ \right)
\]
so the direction of motion cannot suddenly change, unless along one
part of the extremal the velocity
of point $i$ is zero (the point is at rest), where its direction
$\vhat$ is then undefined. 

\begin{figure}
\centering
\begin{tabular}{|cc|cc|} \hline
a&&b& \\
 & \includegraphics[height=0.3\linewidth]{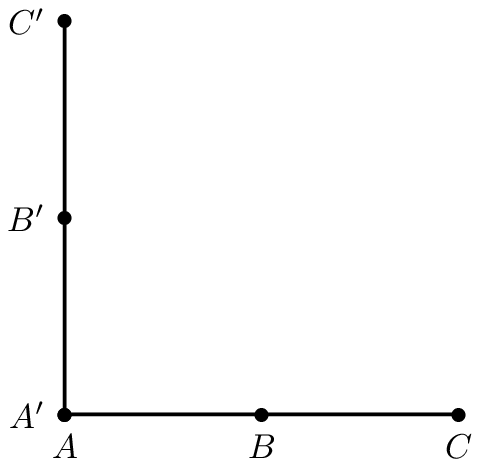} &  &
\epsfig{file=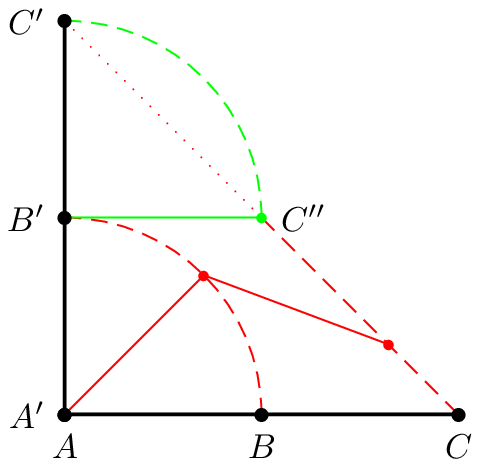,width=0.3\linewidth,clip=}  \\
\hline
c && d & \\
 & \includegraphics[height=0.3\linewidth]{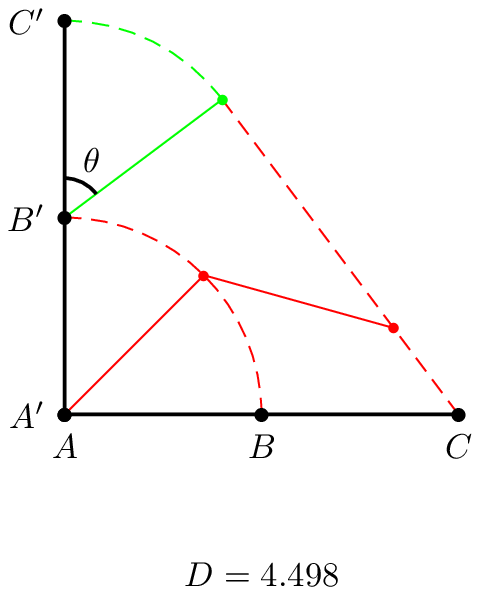} &  & 
\includegraphics[height=0.3\linewidth]{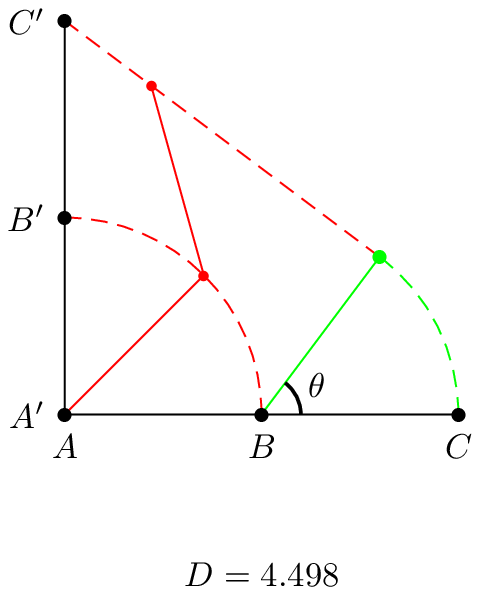} \\
\hline
\end{tabular}
\caption{(a) Initial and final states for a chain of two links. The
  transformation in (b) is non-extremal because it violates a corner
  condition at $C''$. (c) and (d) are degenerate minima- rotations
  occurring about $B'$ or $B$ both have the same length. Intermediate
  states is shown in red have opposite convexity in (c) and (d). } 
\label{figLcurv}
\end{figure}

The boundary conditions described in section~\ref{secBCs} hold as
well, so the end points can either be at rest, move in straight lines,
or purely rotate. This gives $3\times 3=9$ possible scenarios to
investigate here, many of which can readily be ruled out. For example
consider the states in figure~\ref{figLcurv}a. Because $A$ and $A'$ are
in the same position, rotation and translation of $A$ are ruled out and
point $A$ remains at rest,
leaving $3$ scenarios for the other end point $C$. However since $C$
and $C'$ are at different positions and $ABC$ are along a straight
line, $C$ cannot remain at rest initially, leaving either translation or
rotation for point $C$.

Suppose $C$ translates towards $C'$ as in figure~\ref{figLcurv}b.
Then $\vhatdot_{\mbox{\tiny{C}}} =0$ and
from~(\ref{eq:ELeqnsC2link},\ref{eq:ELeqnsB2link})
$\l_{\mbox{\tiny{BC}}}=0$ and $\vhatdot_{\mbox{\tiny{B}}}
=\l_{\mbox{\tiny{AB}}} \, \bfr_{\mbox{\tiny{B/A}}}$. $B$ cannot move
in a straight line without moving point $A$, so
$\l_{\mbox{\tiny{AB}}}\neq 0$ and thus $B$ must rotate about point
$A$. The transformation then proceeds as in figure~\ref{figLcurv}b
until $B$ reaches $B'$ and $C$ reaches $C''$. Then however if $C''$
were to rotate to $C'$, the trajectory would violate corner conditions
at point $C''$. Therefore the direction of translation of $C$ must not
be directly to $C'$ but must be tangential to the arc 
$\stackrel{\frown}{C'C''}$ as in figure~\ref{figLcurv}c. 

The reverse of this transformation is allowable as well, as can be
seen by swapping the labels $ABC\rightarrow A'B'C'$. Here $C$ first
rotates to the critical angle $\th$ shown in fig~\ref{figLcurv}d
and then translates to $C'$. 

\begin{figure}
\centering
\begin{tabular}{ccc} 
\includegraphics[height=0.3\linewidth]{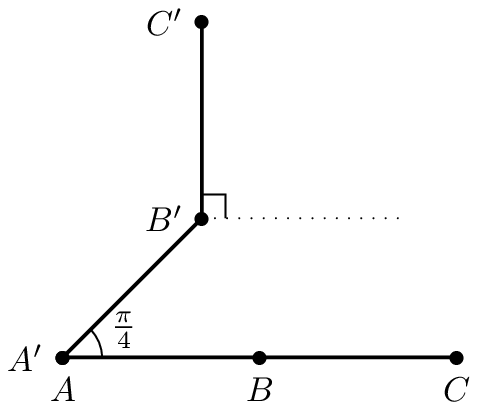}
& \includegraphics[height=0.3\linewidth]{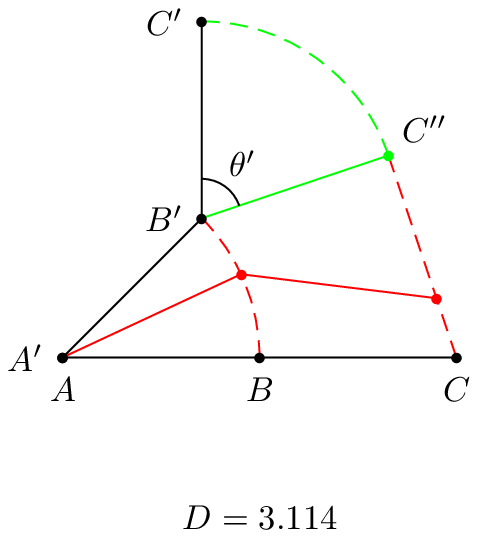}
& \includegraphics[height=0.3\linewidth]{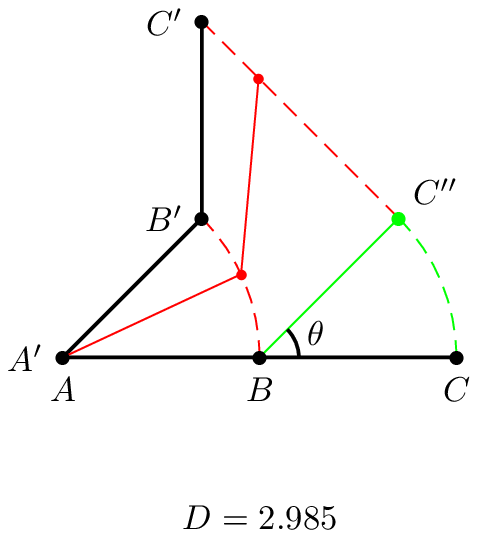}
\\
a & b & c \\
\end{tabular}
\caption{(a) Initial and final states for a polymer of $2$ links. The
  angle between $\ovl{AB}$ and $\ovl{A'B'}$ is $\pi/4$. The minimal
  transformations in (b) and (c) are now no longer degenerate. (c) is
  the global minimum. }
\label{figTT}
\end{figure}

In fact one can see that links $BC$ and $B'C'$ along with lines
$\overline{BB'}$ and $\overline{CC'}$ form a quadrilateral as in
figure~\ref{rotfigs}, with the same consequences for rotation to a
critical angle. For the links in fig~\ref{figLcurv} the situation is
symmetric so rotation can occur at the beginning or end of the
transformation. Figure~\ref{figTT}a shows an example with this
symmetry broken, so that the distance is different depending where the
rotation occurs, as in figures~\ref{rotfigs}a,b. In this case, the
transformation in fig~\ref{figTT}c has the minimal distance, and
that in fig~\ref{figTT}b is subminimal. Extensions of the 
transformation in
figure~\ref{figTT} to large numbers of links were explored
in~\cite{PlotkinSS07}. 

\subsection{Transformations involving a change in convexity}
\label{sec:2linkconvexity}

Transformations between configurations with opposite convexity involve
motion out of the plane, even if the initial and final states lie in
the plane. If the transformation is constrained to lie in plane, the
trajectories of some points will be non-monotonic- those points must
move farther away from their final positions before approaching them.  
We illustrate these ideas with some examples below. 

\begin{figure}
\includegraphics[height=0.4\linewidth]{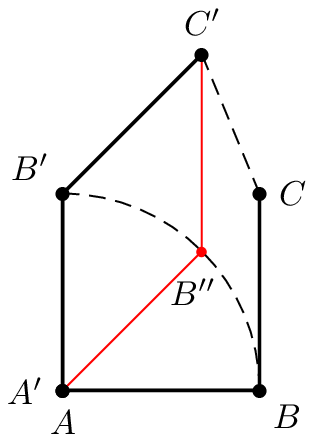}
\caption{A transformation between two states of opposite convexity:
  $ABC$ has convexity down and right, while $A'B'C'$ has convexity up
  and left. There is no extremal
  transformation in the plane that can connect them, without some
  apparent violation of corner conditions. }
\label{figUU}
\end{figure}

Consider the initial and final states in figure~\ref{figUU}. 
We again imagine $B$ rotating to $B'$. If C were to translate to $C'$
one would have the intermediate configuration $A'B''C'$. Now $C'$ and
$A'$ must remain at rest to satisfy corner conditions. Then the only
way to finish the transformation is for $B''$ to rotate about the axis
$\overline{A'C'}$, however then the trajectory of $B$ violates corner
conditions and so is not extremal. In~\ref{app:min2d} we take up
the issue of minimal transformations for this case when the links are
{\it constrained} to lie in a plane. 

We thus seek a point $B''$ and resulting trajectory
$\overrightarrow{BB''B'}$ such that arc $\stackrel{\frown}{BB''}$
satisfies corner conditions with arc $\stackrel{\frown}{B''B'}$. 

\begin{figure}
\centering
\begin{tabular}{cc} 
\includegraphics[height=0.4\linewidth]{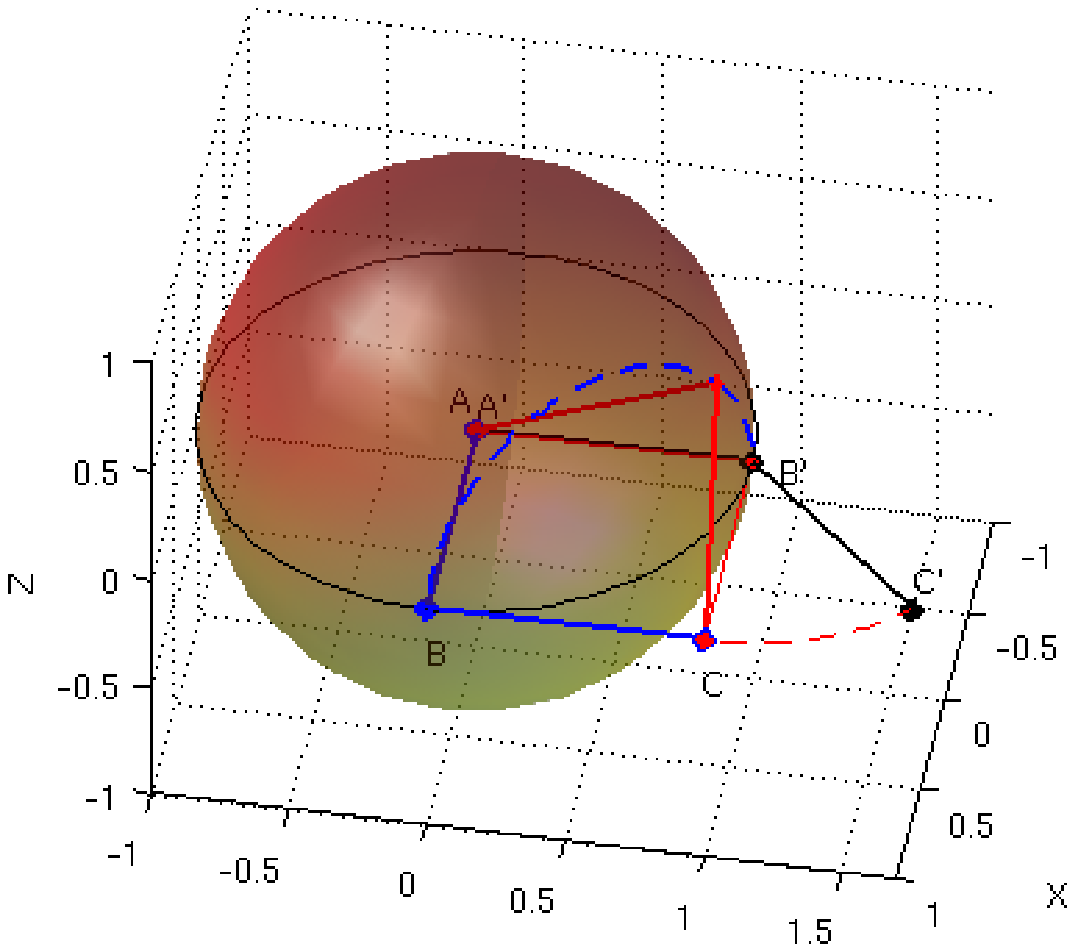}
& \includegraphics[height=0.4\linewidth]{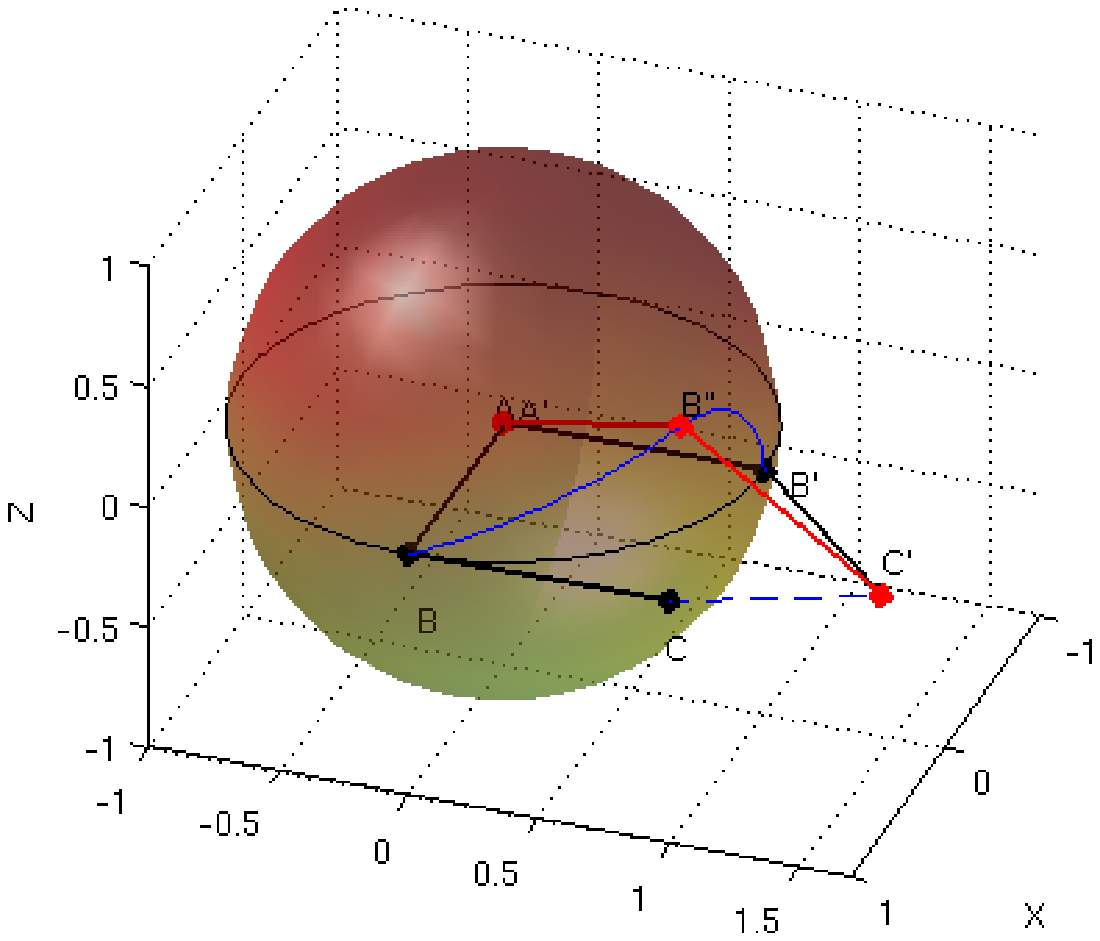}
\\
a & b  \\
\end{tabular}
\caption{Subminimal (a) and minimal (b) transformations for the
  boundary conditions in figure~\ref{figUU}. The distances for each
  transformation are approximately $3.007 L^2$ and $2.576 L^2$
  respectively. Transformation (a) proceeds from $ABC$ by first
  rotating $B$ to $B'$ about axis $\ovl{AC}$, then rotating $C$ about
  point $B'$. Transformation (b) proceeds from $ABC$ by simultaneously
translating $C$ to $C'$ while rotating $B$ about $A$ on a great circle
to point $B''$. Finally point $B$ rotates from $B''$ to $B'$ about
axis $A'C'$. } 
\label{figConvextrans1}
\end{figure}

One solution is to effectively place $B''$ at position $B'$ by
considering the boundary condition with $C$ at rest (and $A$ at
rest). Then $B$ rotates to $B'$ about axis $\overline{AC}$, and the
trajectory of $B$ lies on a circle defined by the intercept of two
unit spheres centered at $A$ and $C$. The sphere about $A$ is drawn in
figure~\ref{figConvextrans1} as a visual aid. Along arc
$\stackrel{\frown}{BB'}$ both $\l_{\mbox{\tiny{AB}}}\neq 0$ and
$\l_{\mbox{\tiny{BC}}}\neq 0$. Once in configuration $A'B'C$, $C$ can
then undergo rotation about $B'$ to $C'$, with $A'$ and $B'$
stationary. 

The transformation in~\ref{figConvextrans1}a is a local minimum in
distance, however it is not the global minimum. A shorter distance
transformation can be seen by considering the reverse
transformation. Imagine $A'$ and $C'$ stationary while $B'$ rotates
about axis $\overline{A'C'}$ in figure~\ref{figConvextrans1}b. This
rotation of $B'$ follows a circular trajectory defined by the intercept of two
unit spheres centered at $A'$ and $C'$. The rotation 
occurs until point $B''$, which is the point where above
circle is tangent to a great circle on the unit sphere about $A$ and
passing through $B$. The arc $\stackrel{\frown}{BB''}$ is a great
circle because this is a geodesic for point $B$ given $A$ is fixed,
which follows from the Euler
equations~(\ref{eq:ELeqnsB2link},\ref{eq:ELeqnsC2link}) when
$\l_{\mbox{\tiny{BC}}} =0$. The great circle is defined by the plane
containing the points $A$, $B$, and $B''$. 

The angle between the (variable) vector $\overrightarrow{BC}$ of link $BC$ and
the tangent the 
the arc $\stackrel{\frown}{B'B''}$ is always $\pi/2$, so once the
corner condition is met, point $C$ on link $BC$ can move in straight
line motion from $C'$ to $C$ while $B$ moves on the great circle from
$B''$ to $B$. That is, the quadrilaterial criterion of
section~\ref{sec:stline} is met for $\Box B B'' C' C$.

To find point $B''$, let its position be $\bfr_{\mbox{\tiny{B''}}}=
(x_o,y(x_o),z(x_o))$. The great circle is defined by the plane passing
through the points $A$, $B$, and $B''$. This plane has normal
$\bfn \equiv \overrightarrow{AB} \times \overrightarrow{AB''} =
(1,0,0) \times (x_o,y(x_o),z(x_o)) = (0,-z(x_o), y(x_o) )$. At the
point $B''$ the normal is orthogonal to the tangent vector of the
circle defined by rotation about the $AC'$ axis. This tangent vector
is $\tangent = \D \bfr /\D s = x_s (1, y_x , z_x)$ by the chain
rule. At $B''$, $\tangent \cdot \bfn = 0$, or 
\begin{equation}
  \label{eqcrosscond}
-z(x_o) y_x (x_o) + y(x_o) z_x(x_o) = 0  
\end{equation}
The functions $y(x)$ and $z(x)$ are defined by the intercept of two unit
spheres centered at $(0,0,0)$ and $(1/\sqrt{2}, 1+1/\sqrt{2}, 0)$,
giving 
\par 
\begin{align}
y(x) &= 1- {\sqrt{2} \over {2+\sqrt{2}}} x \nonumber \\
z(x) &= \sqrt{1-x^2 - y(x)^2} \: .
\label{eqyzofx}
\end{align}
Together~(\ref{eqcrosscond}) and (\ref{eqyzofx}) give 
\par  \begin{align*}
\bfr_{\mbox{\tiny{B''}}}= 
\begin{pmatrix} 
\sqrt{2}-1 \\  
2 ( \sqrt{2}-1 ) \\
\sqrt{2 ( 5\sqrt{2} -7)}
\end{pmatrix}
\end{align*}
The distance travelled along arc $\stackrel{\frown}{BB''}$ is 
$\th_{\mbox{\tiny{BB''}}}$, where
$\cos \th_{\mbox{\tiny{BB''}}} = x_o = \sqrt{2}-1$. 
The distance travelled along arc $\stackrel{\frown}{B''B'}$ can 
similarly be shown to be $r \th_{\mbox{\tiny{B''B'}}} = \sin(\pi/8)
\cos^{-1} ( 2 \sqrt{2}-3)$. Adding the distance $\overline{CC'}$, the
total (minimal) distance is thus $\Dist = 2.576$. There is of course a
degenerate solution to the above with $z\rightarrow -z$. 

\subsection{Transformations with initial and final states in 3-D}
\label{sec:2link3D}

\begin{figure}
\centering
\begin{tabular}{cc} 
\includegraphics[height=0.4\linewidth]{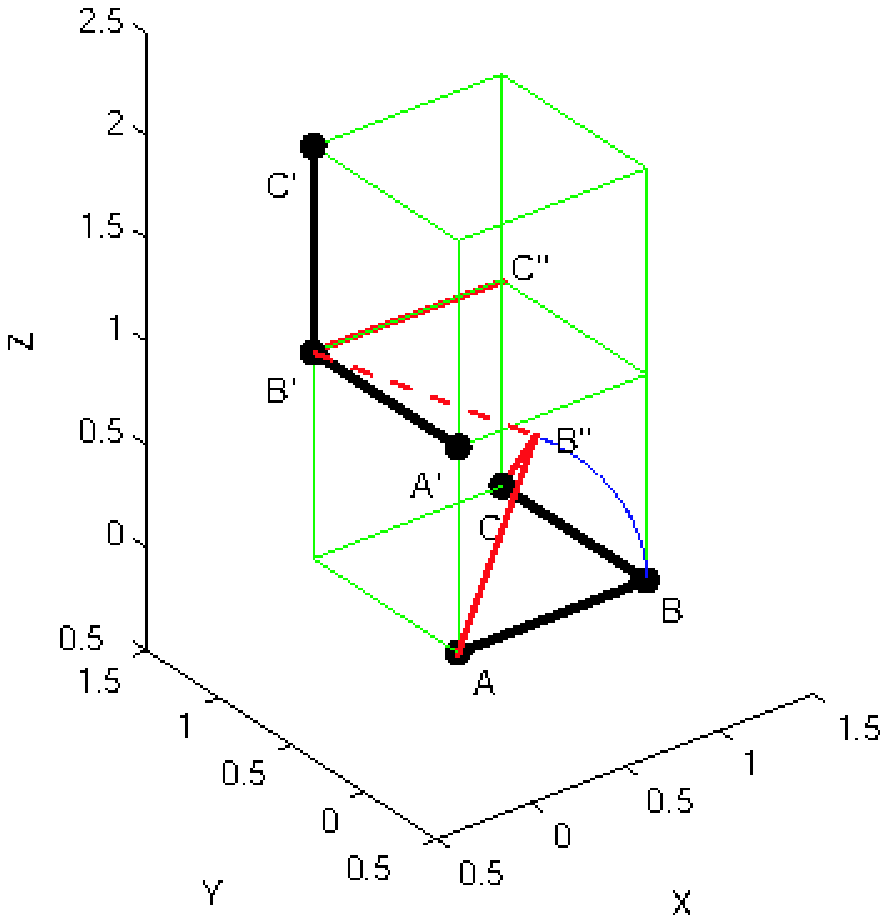}
& \includegraphics[height=0.4\linewidth]{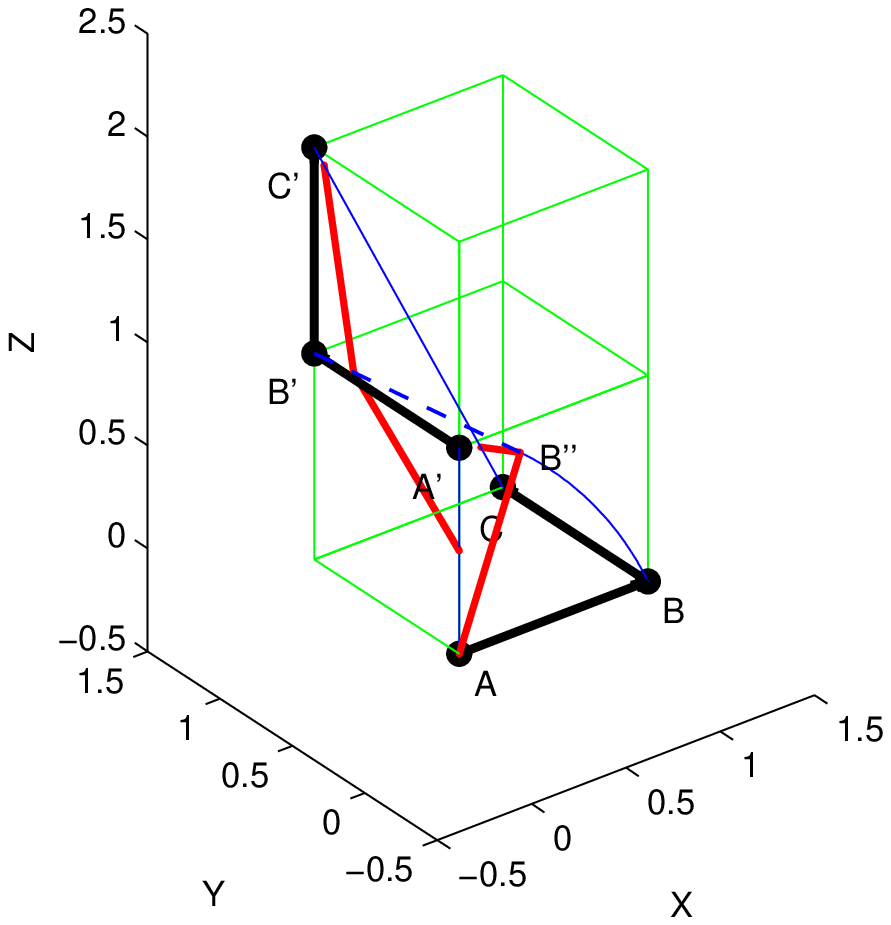}
\\
a & b  \\
\end{tabular}
\caption{(a) Subminimal transformation and (b) minimal
transformations between $ABC$ and $A'B'C'$ (see text)}. 
\label{2linkcube}
\end{figure}

We now give a representative example where the
initial and final configurations do not lie in the same plane, as
shown in figure~\ref{2linkcube}. Because $\ovl{AB} \perp \ovl{AA'}$
and $\ovl{BC} \perp \ovl{CC'}$, neither $A$ nor $C$ will rotate about
$B$ as part of the transformation. Nor can $ABC$ simultaneously
translate directly to $A'B'C'$, because for example quadrilateral
$\Box AA'B'B$ does not satisfy the rule of opposite angles $\geq
\pi/2$, so link $AB$ cannot slide (translate) to $A'B'$. 

This leaves $3$ options for the initial stages of the transformation: \\
{\bf 1.)} $A$ translates, $B$ rotates, $C$ remains
fixed. $B$ then rotates about $C$ in the $CBB'$ plane. The initial
direction of motion of $B$ is then $\vhatB = (-\ihat +
\khat)/\sqrt{2}$, however then $\vhatA$ can only move backward to
preserve link length ($\vhatA = -\khat$), similar to
figure~\ref{figslideproofnec}. This rules out case {\bf (1)}. \\
{\bf 2.)} $A$ remains fixed, $B$ rotates, $C$ remains fixed. $B$ then
rotates towards $B'$ about axis $\ovl{AC}$ until it reaches a critical
angle where line $\ovl{B''B'}$ is tangent to its circular
trajectory (see fig.~\ref{2linkcube}a). 
At this point the quadrilateral $\Box B'' C C' B'$ does
not have opposite obtuse angles, so a straight line transformation to
$A'B'C'$ is not possible. It is possible to transform to a
configuration $A'B'C''$, where $C''$ is at position $(1,1,1)$ and
angle $\angle B'C''C =\pi/2$, so that $\vhatC = \khat$. 
Then the transformation is completed by 
a $\pi/2$ rotation of $C''$ about $B'$. This transformation is
subminimal. \\
{\bf 3.)} $A$ remains fixed, $B$ rotates, $C$ translates.
In this case, $B$ rotates toward $B'$ in the $BAB'$ plane, while $C$
translates to $C'$, until the state $AB''C''$ is reached (see
fig.~\ref{2linkcube}b). State $AB''C''$ can be found as
follows. Because the rotation of $B$ is about the axis
$(0,-1/\sqrt{2},1/\sqrt{2})$, the position $\ora{AB''}$ of $B''$ after
rotation of the (critical) angle $\th$ is
$(\cos\th, \sin\th/\sqrt{2} , \sin\th/\sqrt{2})$. This angle is
then determined by the condition $\ora{AB''} \cdot \ora{B''B'}=0$, where
$\ora{B''B'} = \ora{AB'}-\ora{AB''}$. The solution to this condition
is simply $\th=\pi/4$. The location of $C''$ is then determined from
the condition that the link length from $B''$ to $C''$ is one:
$|\ora{B''C''}| = 1$, where $\ora{B''C''}= \ora{AB''}+ t\ora{CC'}$.
Solving this condition for $t$ gives the position of $C''$ as
$( {3+\sqrt{2} \over 5} , 1, {2 (2-\sqrt{2})\over 5})$.
At this point the quadrilateral $\Box B'B''C''C'$ has opposite
obtuse angles, and quadrilateral 
$\Box AB''B'A'$ has opposite angles
$=\pi/2$, so it is in a bowtie configuration as in the end point
configurations in 
figure~\ref{bowtie}. Therefore 
all points $AB''C''$ can translate from this intermediate state to
their final positions $A'B'C'$. The total distance travelled is $\th +
|AA'| + |CC'| +|B''B'|$ or $\Dist = 2+\pi/4+\sqrt{5}\approx 5.022$. 
The reverse of this transformation is also possible, where point $B'$
rotates about $A'$ in the plane $B'AB$, while $C'$ translates along
$\ora{C'C}$. Inspection reveals the distance covered is the same as the
forward transformation.

\section{Limit of large link number}
\label{sec:largelink}

\begin{figure}
\centering
\begin{tabular}{cc} 
\includegraphics[height=0.3\linewidth]{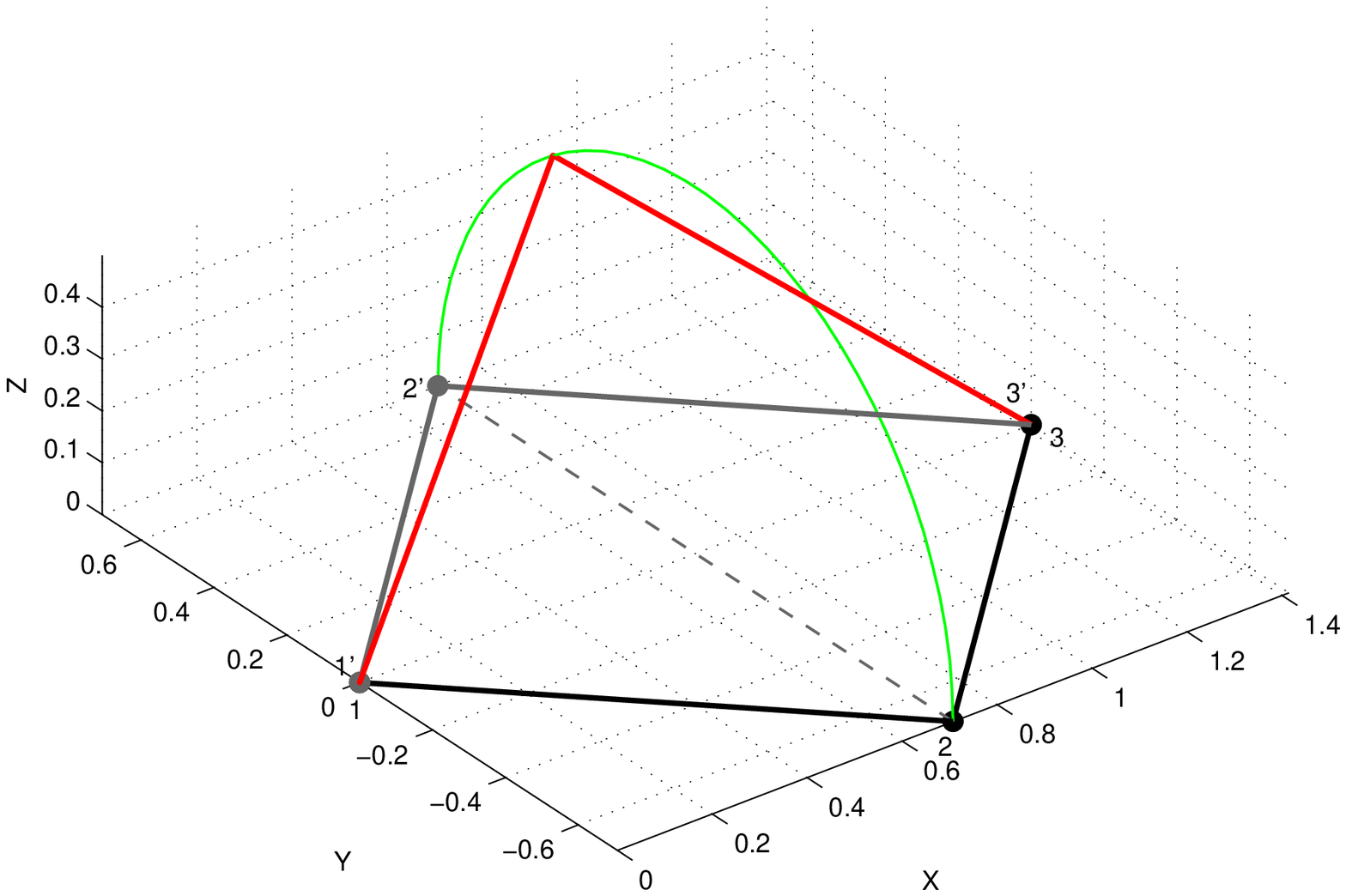}
& \includegraphics[height=0.3\linewidth]{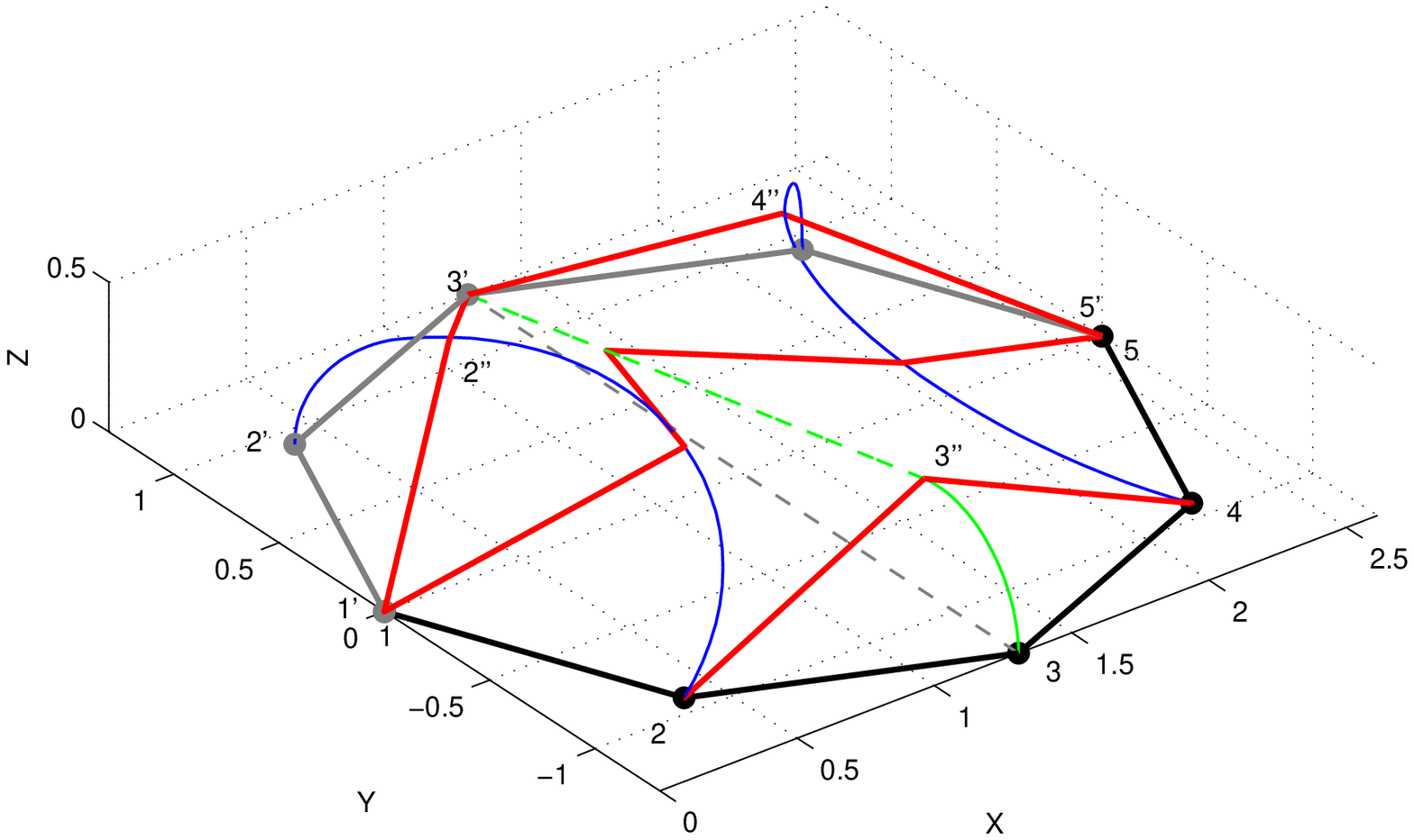}
\\
a & b  \\
\includegraphics[height=0.3\linewidth]{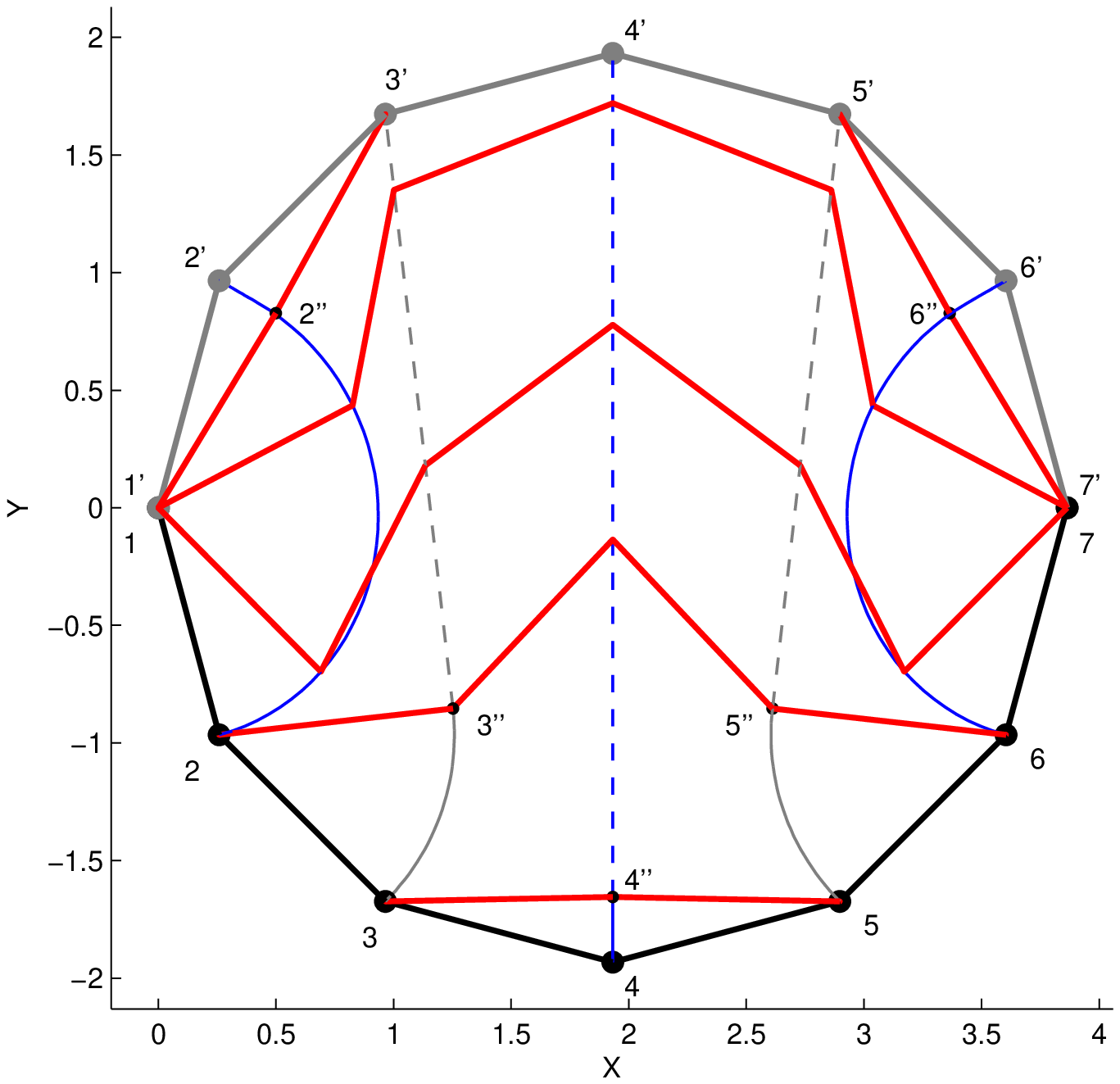}
& \includegraphics[height=0.3\linewidth]{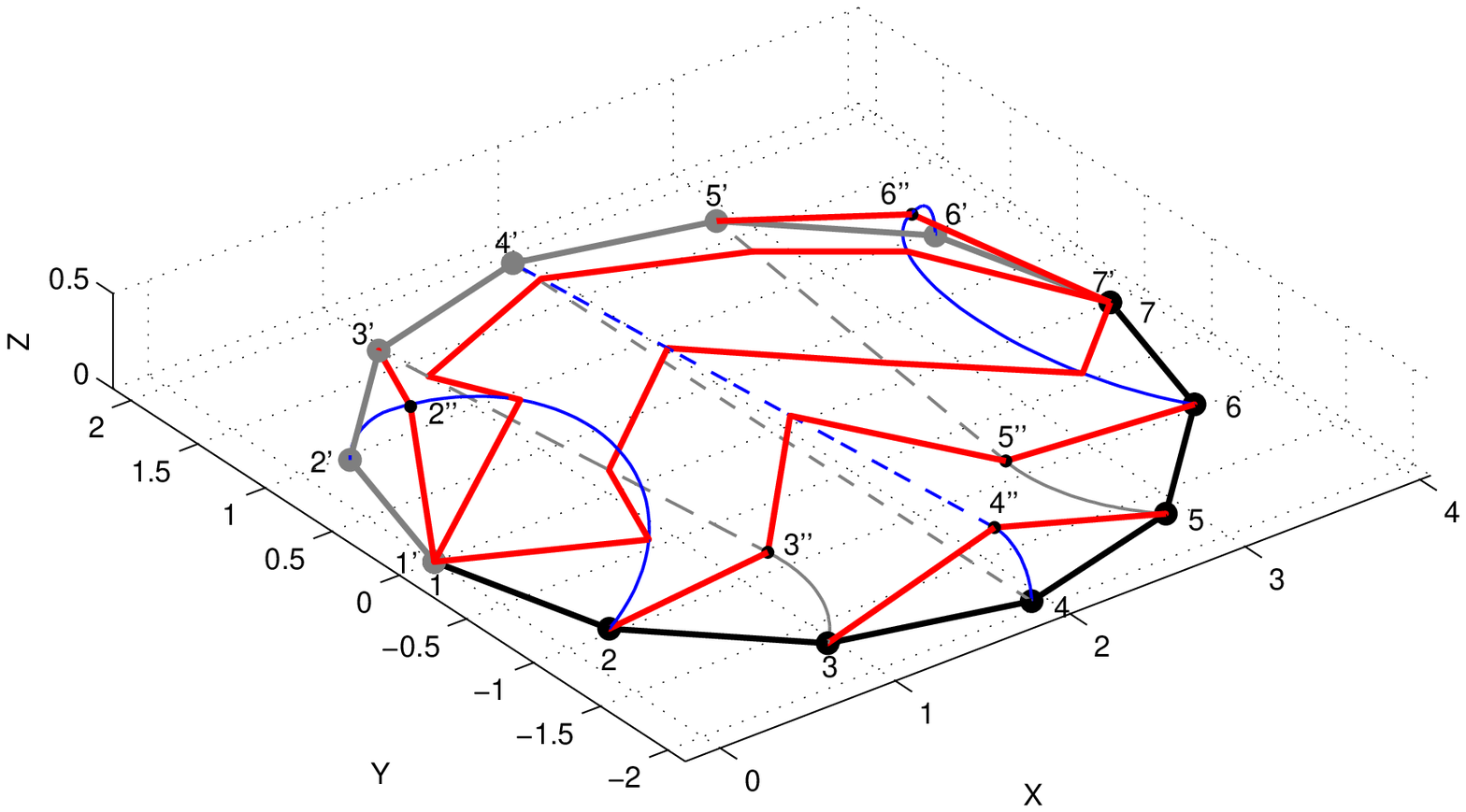} \\
c & d \\
\end{tabular}
\caption{Examples of transformations between initial and final states
  of opposite convexity, for increasing numbers of links. (a)
  illustrates the transformation for $N=2$ links.  (b) $N=4$ and
  initial and final state form an octagon. (c,d) $N=6$ and initial and
  final states form a dodecagon. (c) top view. (d) view in
  perspective. Rotations are shown as solid color lines (either green
  or blue). Translations are shown as dashed lines. The grey dashed
  lines underneath $\ovl{3''3'}$ in (b) and $\ovl{4''4'}$ in (d) are
  shown only to illustrate that those lines are above the plane.}
\label{limlink}
\end{figure}

From the transformation discussed in section~\ref{sec:2linkconvexity},
we see that if both $\angle ABC$ and $\angle A'B'C'$ were $\pi/2$ as
in figure~\ref{limlink}a, then the transformations in
figures~\ref{figConvextrans1}a and~\ref{figConvextrans1}b become
degenerate, having distance $\Dist=\pi/\sqrt{2}$. The transformation is
completed by a single rotation about axis $\ovl{13}$. 

We can now examine the effect of increasing the link number. Let the
number of links increase to $4$, and let us preserve the symmetry that
is present about the horizontal axis in fig~\ref{limlink}a, so the
initial and final states become an octagon (figure~\ref{limlink}b). In
the limit $N\rightarrow \infty$, the figure becomes a circle. 

If we separated the links in figure~\ref{limlink}a by some distance in
the $y$ direction (perpendicular to axis $\ovl{13}$), then the minimal
transformation involves the same rotation of $2$ about axis $\ovl{13}$
until a critical angle $\theta_c$, after which all three points $123$
can translate in straight lines to $1'2'3'$. In the same fashion, the
minimal transformation for the octagonal transformation in
fig~\ref{limlink}b involves a rotation of point $3$ out of the plane
about axis $\ovl{24}$ to a critical angle $\theta_c$ at which the
point is located at position $3''$.  Once this critical angle is
reached, point $3$ translates in a straight line from $3''$ to $3'$.

Because points $1$ and $5$ are stationary to satisfy corner
conditions, points $2$ and $4$ must move in great circles about points
$1$ and $5$. However points $2$ and $4$ cannot finish the
transformation by moving on great circles. At the configuration $1'2''
3' 4'' 5'$ in figure~\ref{limlink}b, point $3$ has finished the
transformation, but points $2$ and $4$ have not. 
To satisfy corner conditions at the points $2''$ and $4''$, the great
circles must be out of plane as well. At points $2''$ and $4''$, the
transformation finishes with rotations about axes $\ovl{1'3'}$ and
$\ovl{3'5'}$. The total distance $\Dist \approx 7.93$.

Of course the time reverse of this transformation (equivalent to
swapping primed and unprimed labels) is also a minimal transformation, as is
the transformation obtained by reflection about the $z=0$ plane. 

Now consider increasing the chain to $6$ links, so the combination of
$\bfr_i(0)$ and $\bfr_i(T)$ becomes a dodecagon (12-sided polygon, see
figures~\ref{limlink}c-d). As
before the midpoint vertex (here $\bfr_4$) must rotate out of the plane
about axis $\ovl{35}$ to a critical angle $\th_c$ before translating
in a straight line to $\bfr_{4'}$. This critical angle is where
$\ora{34''} \cdot \ora{4''4'} = \ora{54''}\cdot\ora{4''4'} =0$. The
quadrilaterals $\Box 2 2' 3' 3$ and $\Box 655' 6'$ are of the type in
figure~\ref{rotfigs}, so point $3$ must rotate about $\bfr_2(0)$ to a
critical angle where $\ora{23''} \cdot \ora{3''3'} =0$, and likewise for
point $5$. 

While point $3$ rotates to its critical angle, point $4$ translates
along line $\ovl{4''4'}$. Points $\bfr_1(0)$ and $\bfr_7(0)$ overlap with
$\bfr_1(T)$ and $\bfr_7(T)$ and so remain fixed to satisfy corner
conditions. After point $3$ has reached its critical angle, it can
translate along $\ovl{3''3'}$ as point $2$ rotates about
$\bfr_1$. However to satisfy corner conditions at point $2''$, the
rotation cannot remain in the $x-y$ plane. Point $\bfr_{2''}$ is
determined as the point where $\tangent\cdot\normal_{plane} =0$, where
$\tangent$ is the tangent to the arc $\stackrel{\frown}{2 2''}$
defined by rotation about axis $\ovl{13'}$, and $\normal_{plane}$ is
the normal to the plane $122''$, i.e. $\bfr_{2/1} \times
\bfr_{2''/1}$. The same process holds for point $6$. These critical
points and some intermediate states for the transformation are shown in
figure~\ref{limlink}d. The total distance covered by the
transformation is $\Dist \approx 16.3$.

It is sensible to consider the total length of chain as fixed to say
$L=1$, and to let the link length $d\sN$ for the chain of $N$ links be
determined by $N d\sN=L$. Because distances scale as $d\sN^2$, the
$N=2,4,6$ cases have $\Dist_2 \approx 0.555 L^2$, $\Dist_4 \approx
0.496 L^2$,
$\Dist_6 \approx 0.445 L^2$. Note that this distance decreases with
increasing number of links: the constraints on the 
motion of the various beads during the transformation are 
relaxed as the number of links is increased. 

We can then imagine resting a piece of string on a table in
the shape of a semi-circular arc, and then asking how one can move
this string to a facing semicircle of opposite convexity. So long as
the string has some non-zero persistence length $\lp$, the
transformation of minimal distance must involve lifting the string off
of the table to change its local convexity. The 
vertical height the string must be lifted (see fig~\ref{limlink}d) is
of order $\sim  \sin (\pi\lp/L) \sim \lp/L$, which goes to zero for an
infinitely long chain. 

As the number of links $N \rightarrow \infty$, some simplifications
emerge. In particular 
the contribution to the total distance due to rotations becomes
negligible, and the translational component dominates. To see this
note that the distance due to straight line motion scales as:
\[
\Dist (\mbox{st. line}) \sim ds \, N L \sim L^2
\]
while the distance travelled during rotations scales as
\[
\Dist (\mbox{rot.}) \sim ds \, N (\ovl{\th_c} ds) \sim L^2/N
\]
where we assume the worst case scenario where an extensive number of
links must rotate before translating. 
Because translation dominates the distance as $N\rightarrow \infty$,
the distance travelled converges to $L$ times the mean root square
distance (MRSD), i.e.
\begin{align}
\Dist_{\infty} &\rightarrow ds \, \sum_{i=1}^{N+1} \left| \bfr_i(T) - \bfr_i(0)
\right|  \nonumber \\
&= L {1 \over N} \sum_i \sqrt{ \left(\rBi -\rAi \right)^2} \nonumber \\
& = L \left( \mbox{MRSD} \right) 
\label{eq:mrsd}
\end{align}

The MRSD for the examples in figures~\ref{limlink}b,d are $0.394\,L$ and 
$0.400\, L$ respectively, which are both less than the actual
distances travelled (in units of $L$). In the limit $N\rightarrow
\infty$, where the polygon becomes a circle, the distance converges to
$\Dist_\infty = 4 L^2/\pi^2 \approx 0.4053 L^2$.  
For large $N$ systems then, it is a good first approximation to use
$MRSD$ for the distance. 

The MRSD is always less than the root mean square distance (RMSD),
except in special cases when they are equal. 
To see this, we can apply H\"{o}lder's inequality
\[
\sum_{k=1}^N \left(g_k\right)^\alpha \,  \left(h_k\right)^\beta \leq
\left(   \sum_{k=1}^N g_k 
\right)^\alpha \left( \sum_{k=1}^N h_k \right)^\beta 
\]
where $g_k , h_k \geq 0$, $\alpha, \beta \geq 0$, and  $\alpha +\beta
=1$.  With the specific identifications $g_k = (\rBk -\rAk )^2 \equiv
\Delta \bfr_k^2$, $h_k=1$, and $\a =\b =1/2$, we have directly
\[
{1\over N} \sum_{k} \sqrt{\Delta \bfr_k^2} \: \leq \sqrt{{1\over N} 
 \sum_{k} \Delta \bfr_k^2 }
\]
For example the RMSD for the circle configuration discussed above is 
$\sqrt{2} L/\pi \approx 0.4502 L$, which is greater than the
MRSD. 

The fact that the distance converges for large $N$ to MRSD rather than
RMSD suggests that RMSD may not be the best metric for determining
similarity between molecular structures, although it is ubiquitously
used.  This fact warrants future investigation- it has implications in
research areas from structural 
alignment based pharmacophore
identification~\cite{GreeneJ94,LemmenC00,PatelY02} to protein
structure and function prediction~\cite{GersteinM98,BakerD01}.

It was shown in~\cite{PlotkinSS07} that chains with persistence
length characterized by some radius of curvature $R$ have extensive
corrections to the MRSD-derived minimal distance, which do not vanish as
$N\rightarrow\infty$, but remain so long as $R/L$ is
nonzero. Likewise, chains that cannot cross themselves have nonlocal
EL equations and extensive corrections to the minimal
distance. Nevertheless, it is worthwhile to investigate some more complex
polymers with MRSD as an approximate distance metric. We pursue this
in the next section. 

\subsection{MRSD as a metric for protein folding}
\label{secprot}

Here we examine the use of MRSD as a metric or order parameter
for protein folding. To this end we adopt an unfrustrated $C_\a$ model
of segment $84-140$ of {\it src tyrosine-protein kinease} (src-SH3),
by applying a G\=o-like Hamiltonian~\cite{Ueda75,Shea2001,ClementiC04}
to an off--lattice 
coarse-grained representation of the {\it src-SH3} native structure
(pdb 1fmk). Amino acids are represented as single beads centered at
their $C_\a$ positions. 
The G\=o-like energy of a protein configuration $\a$
is given by the following Hamiltonian, which we will explain term by
term:
\bea
\Ham(\a | N)
&=& k_r  \sum_{bonds} \left( r_\a - r_N \right)^2 +
k_{\theta}  \sum_{triples} \left( \theta_\a - \theta_N \right)^2  
\nonumber \\
 &+& \sum_{n=1,3} k_{\phi}^{\left(n\right)} \sum_{quads}  
\left[1 - \cos \left( n \times (\phi_\a -
\phi_N) \right) \right]  \nonumber \\
 &+& \en \sum_{j \geq i+3}   \left[ 6 \left(
\frac{\sigma_{ij}}{r_{ij}}\right)^{10}  - 5 \left(
\frac{\sigma_{ij}}{r_{ij}} \right)^{12} \right] + 
\enn \sum_{j\geq i+3} \left(
\frac{\sigma_{ij}}{r_{ij}} \right)^{12}  \: .
\label{go_ham}
\eea
Adjacent beads are strung together into a polymer through harmonic
bond interactions that preserve native bond distances between
consecutive $C_\a$ residues. Here  
$r_\a$ and $r_N$ represent the distances
between two subsequent residues in configurations
$\a$ and the native state $N$. As with other parameters in the
Hamiltonian, the distances $r_N$ are based on the pdb structure and may
vary pair to pair. 
The angles $\theta_N$ represent the angles formed by
three subsequent $C_\a$ residues in the pdb structure, and the angles
$\phi_N$ represent the dihedral angles defined by four
subsequent residues. 
The dihedral potential consists of a sum of two terms, one with period
$2\pi$ and another with $2\pi/3$, which give {\it cis} and {\it trans}
conformations for angles between successive planes of three amino
acids, with a global dihedral potential minimum at $\phi_N \in [-\pi,
\pi]$.  

The parameters $k_r$, $k_{\theta}$, and $k_{\phi}$, are taken to
accurately describe the energetics of the protein backbone: 
we used the
values $k_r= 50 \: \mbox{kcal/mol}$, 
$k_{\theta} = 20 \: \mbox{kcal/mol}$, 
$k_{\phi}^{(1)} = 1 \: \mbox{kcal/mol}$ and 
$k_{\phi}^{(3)} = 0.5 \: \mbox{kcal/mol}$ 
for molecular dynamics (MD) simulations using the AMBER software package.
For MD simulations using LAMMPS, we had used slightly
different values: $k_r= 80\: \mbox{kcal/mol}$, 
$k_{\theta} = 16 \:\mbox{kcal/mol}$, 
$k_{\phi}^{(1)} = 0.8 \: \mbox{kcal/mol}$ and
$k_{\phi}^{(3)} = 0.4 \: \mbox{kcal/mol}$. 

The last line in equation~(\ref{go_ham}) deals with 
non-local interactions, both native and non-native. 
If two amino acids are separated by $3$ more along the chain ($|i-j|\geq
3$), and have one or more pairs of heavy atoms within a
cut-off distance of $r_c = 4.8$ \AA\ in the pdb structure, the amino
acids are said to have a native contact. Then the respective 
coarse-grained $C_\a$
residues are given a Lennard-Jones-like 10-12 potential of depth
$\en = -0.6\: \mbox{kcal/mol}$ 
($-0.8 \: \mbox{kcal/mol}$ for LAMMPS simulations)
and a position of the potential minimum equal
to the distance of the $C_\alpha$ atoms in the pdb structure. 
That is, $\sigma_{ij}$ is taken equal to native distance between $C_\a$
residues $i$ and $j$ if $i$--$j$ have a native contact. 

If two amino acids are not in contact, their respective $C_\a$
residues sterically repel each other ($\enn = +0.6\: \mbox{kcal/mol}$). 
Thus $\enn =0$ if $i$-$j$ is a native residue pair, while $\en=0$ if
$i$-$j$ is a non-native pair. For non-native residue pairs,
$\sigma_{ij} = 4 \: \mbox{Angstroms}$.

In an {\it arbitrary} configuration $\a$, two $C_\a$ residues $i$ and $j$
are considered to have formed a native contact if they have a distance
$r_{ij} \leq 1.2  
\sigma_{ij}$. The results do not strongly depend on the specific value
of this cutoff. The fraction of native contacts present in the 
particular configuration $\alpha$ is then defined as $Q$ (or $Q_\alpha$).

The MRSD of configuration $\alpha$ is found by aligning this
configuration to the native structure, by minimizing MRSD over $3$
translational and $3$ rotational degrees of freedom.  

Constant temperature molecular dynamics simulations were run for this
system using both AMBER and LAMMPS simulation packages. The
probability for the system to have given values of $Q$ and $MRSD$
within $(Q,Q+\Delta Q)$ and $(MRSD, MRSD+\Delta MRSD)$ is proportional to
the exponential of the free energy $F(Q,MRSD)$. Thus the free energy
can be directly obtained by sampling, binning, and taking the logarithm: 
\begin{eqnarray}
  \label{eq:pandf}
  F(Q_1,MRSD_1) - F(Q_2,MRSD_2) = - \kB T \log \left( {p(Q_1,MRSD_1)
      \over p(Q_2,MRSD_2)} \right)
\end{eqnarray}
with $F(1,0) = \En$, the energy of the native structure. 

\begin{figure}
\centering
\begin{tabular}{l} 
a \\
\includegraphics[height=0.56\linewidth]{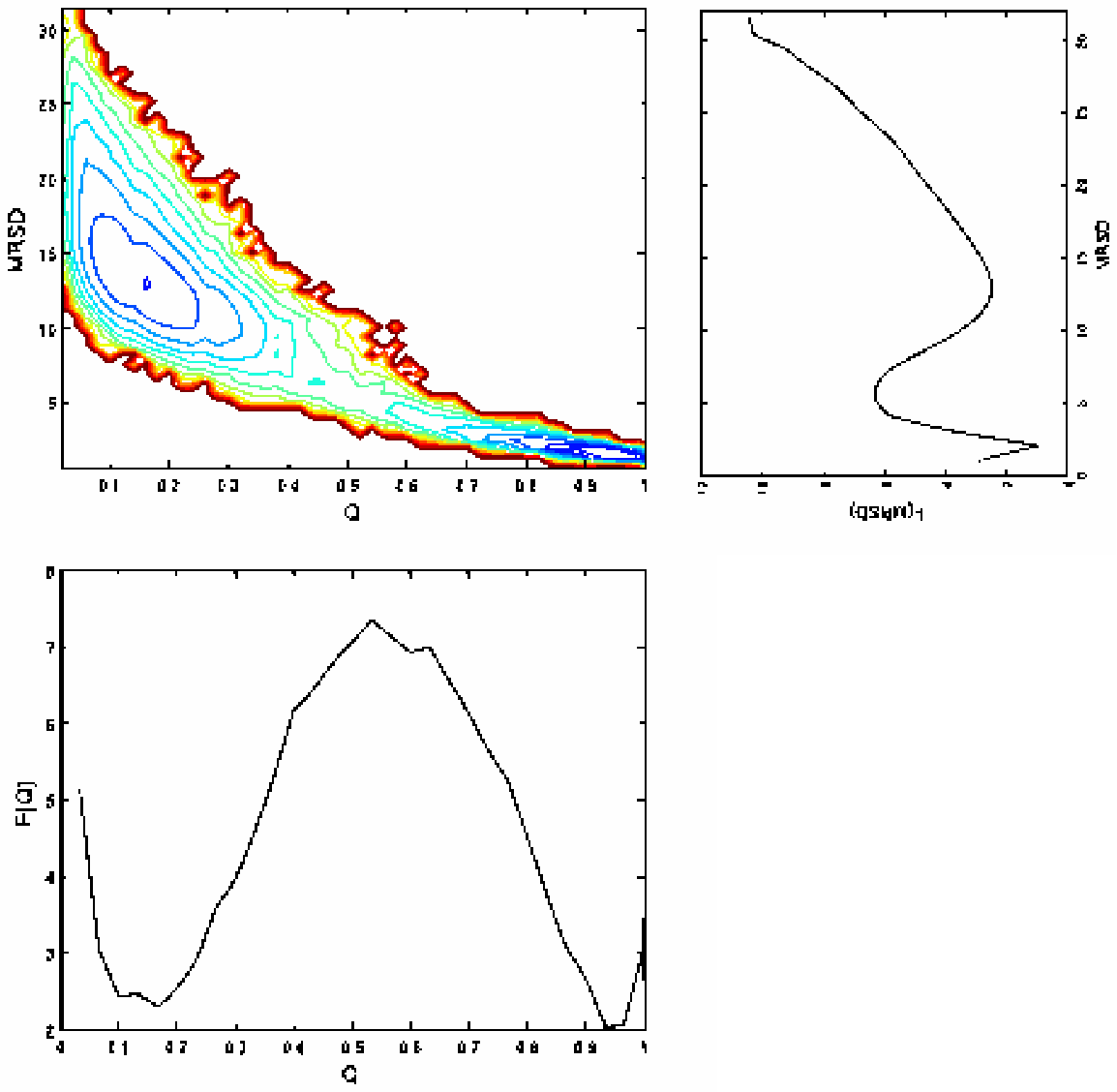} \\
b \\
\includegraphics[height=0.56\linewidth]{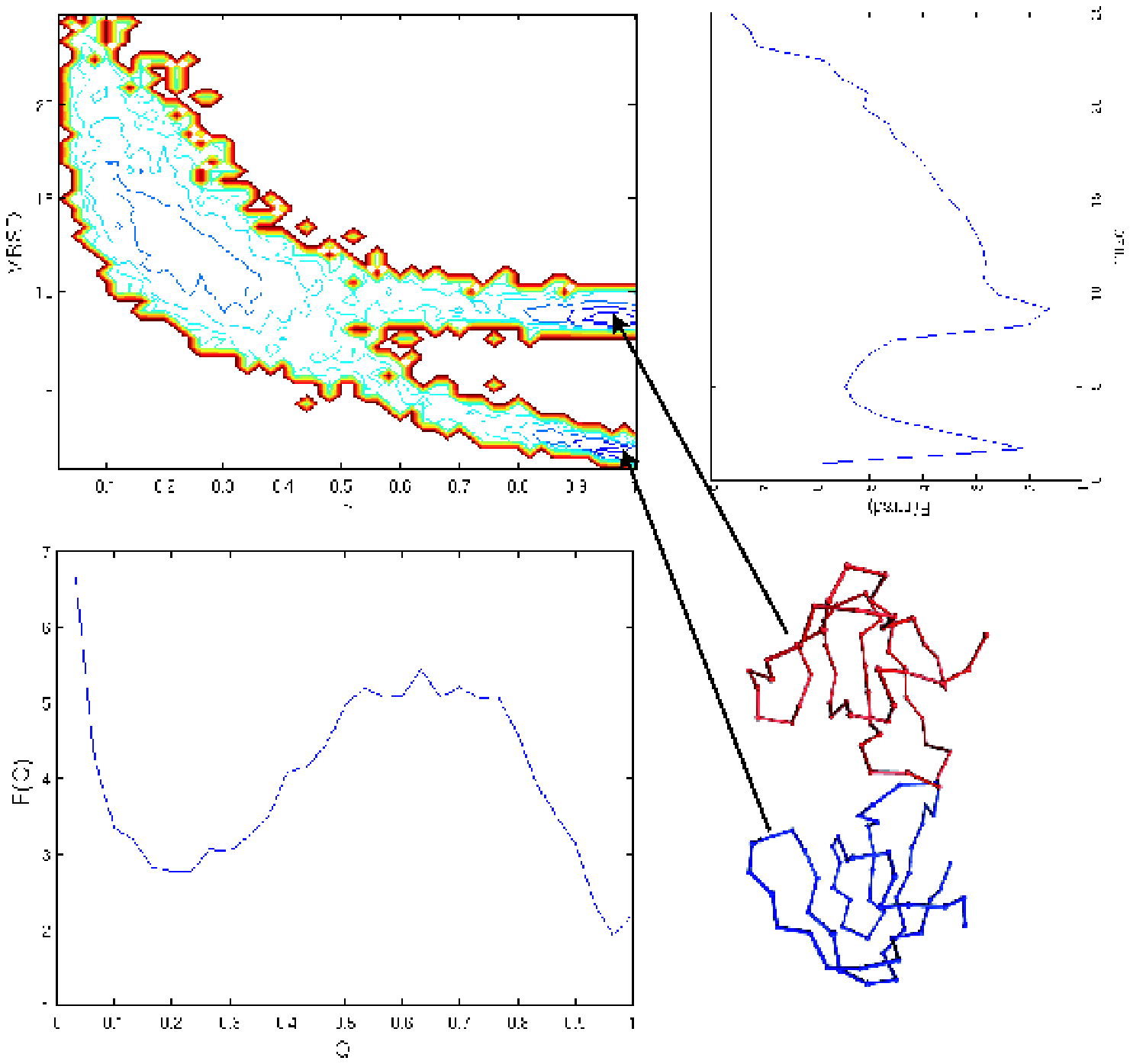} \\
\end{tabular}
\caption{Free energy surfaces for the folding of G\={o}-model {\it
    src-SH3} using 
  two molecular dynamics simulation packages, AMBER (a) and LAMMPS
  (b). The contour plots give $F(Q,MRSD)$. The projections $F(Q)$ and
  $F(MRSD)$ are also shown on each side. The COMPASS class 2 dihedral
  potential in LAMMPS allows for a mirror image of the folded
  structure (red color structure in inset) that is not immediately
  evident from the 
  $F(Q)$ or $F(MRSD)$ surfaces. Future implementations of LAMMPS using
  COMPASS dihedrals for
  biomolecular simulations must then correct for dihedral angles
  defined on the interval $[-\pi,\pi]$. 
}
\label{contourprots}
\end{figure}

Figure~\ref{contourprots} shows the free energy surfaces obtained
using the above recipe, for the AMBER (fig~\ref{contourprots}a) and LAMMPS
(fig~\ref{contourprots}b) molecular dynamics routines.  
The temperature is taken to be the transition or folding temperature
$\Tf$, where the unfolded and folded free energies are equal.

Notice that
$F(Q)$ is comparable for both as it should be, moreover $F(MRSD)$ is
as well.  However the free energy surface plotted as a function of
both $Q$ and $MRSD$ shows a marked difference. In addition to a
native minimum, the LAMMPS routine has an additional minimum at
$Q\approx 0.95$ and $MRSD \approx 8.4$. 
The conformational states in this bin are closely related, with an
average MRSD between them of $1.8$\AA.
We can take the most representative state in this bin as that which
has a minimum 
MRSD from all the others in the bin (at $Q\approx .95$, $MRSD\approx
8.4$): $\minoveri \left( \sum_{j\neq i}^{'} 
  MRSD_{ij}/\sum_{j\neq i}^{'} \right) \approx 1.6$\AA.  
Inspection reveals that this state
is a mirror image of the pdb structure (see
fig~\ref{contourprots}b): If we reflect this structure about one
plane, and subsequently align this reflected structure to the pdb one,
the MRSD is only $1.1$\AA. 

The discrepancy in free energy surfaces corresponding to the presence
of a low energy mirror-image structure arises because
the COMPASS class 2 dihedral
  potentials in the LAMMPS algorithm do not ascribe a sign to the angle 
$\phi$, so the full range 
$[-\pi,\pi]$, is projected onto $[0,\pi]$. This gives the set of
actual dihedral 
angles $\{ \phi_i +\pi \}$ the same energy as the set $\{ \phi_i \}$,
so that the dihedral potentials have two minima rather than
one, and thus a protein chain of the opposite chirality (a mirror image) is
allowed and has the same energy as the pdb structure. We found that
the CHARMM and harmonic dihedral styles do not have this problem,
however they have less versatile function forms, so that we favored
modifying the COMPASS dihedrals to define $\phi$ over its full range.

\section{Conclusions}

Analogously to the distance between two points, the distance between two 
finite length space curves is a variational problem, and may
be calculated by minimizing a functional of $2$ independent
variables $s$ and $t$, where $s$ is the arc-length along the chain,
and $t$ is the 'elapsed time' during the transformation. 

We derived the Euler-Lagrange (EL) equation giving the solution to this
problem, which is a vector partial differential equation, with
extremal solution $\bfr^{\ast} (s,t)$. We also
derived the sufficient conditions for the extremal solution to be a
minimum, through the Jacobi equation. Once the minimal transformation
$\bfr^{\ast} (s,t)$ is known, the distance $\Dist^\ast \equiv
\Dist[\bfr^{\ast}]$ follows. 

We provided a general recipe for the solution to the EL equation using
the method of lines. The resulting $N+1$ EL equations for the
discretized chain are ODEs that can be interpreted geometrically and
solved for minimal solutions. Solutions consist generally of rotations
and translations pieced together so the direction of velocity of any
link end point does not suddenly change (the Weierstrass-Erdmann
corner conditions). 

We explored the minimal transformations for the simplest polymers,
consisting of $1$ or $2$ links, in depth. For transformations between
$2$ links, convexity becomes an issue (the analog to the direction of
the radius of curvature for a continuous string). For example, even if
the initial and final states lie in the same plane, if the convexities
of these states are of opposite sign the transformation must pass
through intermediate states that are out of the plane.
Similarly, given a semicircular piece of string lying on a table, to
move it to a semicircle of opposite convexity using the minimal amount
of motion, the string must be lifted off the table. 

The study of minimal transformations between small numbers of links
has applications to the inverse kinematic problem in robotics and
movement control. In the inverse kinematic problem, one is given the
initial and final positions of the end-effector (the hand of the
robot), and asked for the functional form of the joint variables for
all intermediate states. Generally there is no unique solution until
some optimization functional is introduced, such as minimizing the
time rate of change of acceleration (the jerk),
torque, or muscle
tension (see the review~\cite{KawatoM96} and references therein). The
minimal distance transformation would be 
relevant if one sought the fastest transformation between initial and
final states, without explicit regard to mechanical limitations. 
The indeterminate intermediate points can be handled
variationally as a free boundary value problem. 

In the limit of a large number of links, some simplifications
emerge. For chains without curvature or non-crossing constraints, the
distance converges to $L$ times the mean root square distance
($MRSD$) of the initial and final conformations. 
So for example the distance between 2 strings of length $L$ forming
the top and 
bottom halves of a circle respectively is $4 L^2/\pi^2$, the distance
between horizontal and vertical straight lines of length $L$ which
touch at one end is
$L^2/\sqrt{2}$, and the distance to fold a straight line upon itself
(to form a hairpin) is $L^2/4$. 

The fact that for large $N$ the distance (over $L$) converges to
MRSD rather than 
RMSD suggests that RMSD may not be the best metric for determining
similarity between molecular structures, although it is ubiquitously
used. Adopting MRSD may lead to improvements in structural alignment
algorithms. 

The MRSD was investigated as an approximate metric for protein
folding. Free energy surfaces for folding were constructed for two
simulation packages, AMBER and LAMMPS. It was
found that including MRSD as an order parameter uncovered
discrepancies between the two molecular dynamics algorithms. Because
dihedral angles in LAMMPS (at least in COMPASS class 2 style) are only
defined on $[0,\pi]$, the potential admits a mirror image structure
degenerate in energy with the native structure. This is easily
remedied and should not be interpreted as a deficiency in the LAMMPS
simulation package so long as one is aware of it. It should be
mentioned that the mirror-image structure would also have been seen
had RMSD been used as an additional order parameter. 

It will be important for future studies to address the effects of
persistence length and non-crossing on the distance between biopolymer
conformations~\cite{PlotkinSS07}. Also important is the role of
entropy of paths or transformations in describing the accessibility of
a particular biomolecular structure. Along these lines it will be
interesting to investigate whether the distance can be a predictor of
folding kinetics, or proximity to the native structure. 

It is also an interesting question to ask whether the actual dynamics
between polymer configuraitons resembles the minimal transformation,
after a suitable averaging over trajectories. This question is linked
with the role of the entropy of transformations described above. It is
also related to the problem of finding the dominant pathway for a
chemical reaction~\cite{OnsagerL53}, which has recently been applied
to the problem of protein folding~\cite{SegaM07}. We have focused here
on the question of geometrical distance for complex systems, which can
be separated from the calculation of quantities such as reaction paths
that depend intrinsically on energetics, i.e. on the specific
Hamiltonian of the system. Quantifying the relationship between
geometrical distance and the dominant reaction path is an interesting
future question worthy of investigation. 

The notion of distance and corresponding optimal transformation for a
system with many degrees of freedom is fundamental to a diverse array
of research subjects. Hence we saw potential applications for this
metric in areas ranging from drug design to robotics. It is not clear at
present how useful the calculation of the true Euclidean distance between
high-dimensional 
objects will be for practical applications, but we are optimistic.

\section{Acknowledgements}

We are thankful to Shirin Hadizadeh, Mike Prentiss, and Peter Wolynes
for helpful discussions.  
S.S.P. gratefully acknowledges support from the Natural Sciences and
Engineering Research Council and the A. P. Sloan Foundation. 

\renewcommand{\theequation}{\thesection.\arabic{equation}}
\renewcommand{\thesection}{\Alph{section}}
\setcounter{section}{0}

\appendix

\section{Necessary conditions for straight line transformations}
\label{app:necessary}

It was shown in section~\ref{sec:stline} that to have straight line
transformations between links, it is sufficient to have facing obtuse
angles on opposite sides of the the quadrilateral defined by the
transformation as shown in figure~\ref{obtuse-impossible}A.
We now show that it is a necessary condition as well, i.e. we
show that a slide in the correct direction is not possible in the
absence of obtuse angles.

\begin{figure}[h]
  \centering
  \includegraphics{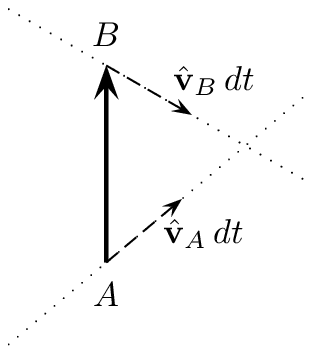}
\caption{}
\label{figslideproofnec}
\end{figure}

Without loss of generality assume that the link is initially along the
z axis.  The paths travelled by the link ends are shown in the figure.
Note that the end point trajectories of $A$ and $B$ are in 3D space so
the paths travelled by $A$ and $B$ need not cross or lie in the same plane.
Let the unit vector along $A$'s path be $\vhatA$ and the unit vector
along $B$'s path be $\vhatB$. Because the
angles that the path of A and the path of B make with the link are
acute, the
z-component of $\vhatB$ ($\equiv \zB$) is negative and the z-component of
$\vhatA$ ($\zA$) is positive. One can write $\vhatA$ and
$\vhatB$ as
\begin{eqnarray}
  \vhatA &=& \rhoA + \zA \zhat \nonumber \\
  \vhatB &=& \rhoB + \zB \zhat \nonumber
\end{eqnarray}
where $\rhoA$ and $\rhoB$ are vectors in xy plane and $\zA > 0$
and $\zB < 0$. 

Let $\rA(t)$ and $\rB(t)$ denote the positions
of the A and B ends at time $t$:
\begin{eqnarray}
  \rA &=& t \vhatA \nonumber \\
  \rB &=& g(t) \vhatB + \zhat \nonumber
\end{eqnarray}
The rigid link constraint dictates that
\[
(\rA - \rB)\cdot(\rA - \rB)=1
\]
which translates to:
\[
g^2 + 2g\left(\zB -t\, (c+\zA\,\zB) \right)-2t\zA+t^2+1=1
\]
with  $c = \rhoA\cdot \rhoB$. 
Solving for $g$ as a function of $t$, keeping in mind that $g(0)=0$:
\[
g(t) = -\left(\zB -t\,\left(c+\zA\,\zB\right)\right) + \sqrt{\left(\zB
  -t\,\left(c+\zA\,\zB\right) \right)^2 - t^2 + 2t\zA} \: .
\]

Now if $g^{\prime}(t) > 0$ it means that the B-end of the link is
travelling in the assumed direction, and if $g^{\prime}(t) < 0$ it means that
B-end is travelling in the opposite direction (which means that the
angle is not acute anymore).
Writing $g^{\prime}(0)$ we get:
\[
g^{\prime}(0) = {{2\,\zB\,c+2\,\zA\,\zB^2-2\,\zA}\over{2\,\left|
      \zB\right| }}+c+\zA\zB = {-\zA \over |\zB|}<0 \: .
\]

Thus point $B$ can only travel in the opposite direction from what was
assumed, which in
turn means an all-acute slide is not possible. We conclude that the
condition of  ``facing obtuse
angles'' is necessary and sufficient for transformations consisting
only of pure translations.

\section{Critical angles}
\label{app:critangles}

The concept of critical angle was first introduced in~\ref{sec:piecewise}. In
order for a straight-line slide of both ends to be possible, at some
stage during the transformation the link needs to rotate about one of
the ends, with the other end being stationary. In principle the
rotation can be about either of the two ends and it can happen at the
beginning or the end of the transformation.
The conditions on the critical angle or orientation can be readily
derived from the broken extremal conditions. It was seen
from~\ref{eq:cornercond1} and~\ref{eq:conjmom}, the non-trivial corner 
conditions read: 
\be
\hat{\bf v}_i|_{{}_+} = \hat{\bf v}_i|_{{}_-} \: .
\label{vhatcorn}
\ee

We know that the path travelled by the moving bead during the rotation
is circular and the path that is travelled during the slide part is a
straight line. Broken extremal condition forces these two paths to be
patched smoothly, which means that the straight-line path should be
tangent to the circle. In the  3D case, for the broken extremal condition to be
satisfied, the straight line slide path and the circular rotation path
should lie in the same plane. For example in figure~\ref{rotfigs} where
$B$ is rotating about 
$A$ initially to $B_1$ and then slides to $B'$, the rotation has to
be in the plane formed by the three points $A B B^{\prime}$. 

Matching the directions of velocity as in~(\ref{vhatcorn}) does not
itself mean that a link can subsequently slide in a straight line,
however at the tangent point, the tangent line to the circle is
perpendicular to the radius, hence one satisfies this second condition
as well.
Below we derive an analytical expression for the critical angle for a
particular case of single link problem, as an example and illustration
of the discussed concepts. Furthermore the particular example will be
used later in~\ref{app:min2d} to introduce minimal
transformations in $2$ dimensions. 

\begin{figure}
  \centering
  \includegraphics{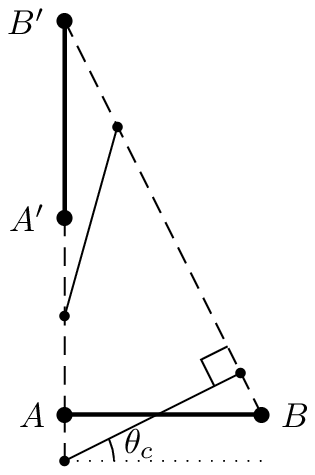}
  \caption{Transformation in which both ends stay on a linear track}
  \label{fig:SingleLinkLinearTrack}
\end{figure}

Consider the single link action with the particular parametrization
$s=s(\theta)$, as discussed in section~\ref{sec:piecewise}:
\begin{equation}
\label{eq:sthetaaction}
 \int(\sqrt{\dot{s}^2 + 1 + 2\dot{s}cos\theta} + \sqrt{\dot{s}^2})\ d\theta.
\end{equation}
where $s \equiv \ora{A(\th)A}$ is the (signed) distance of $A$-end from its initial
position, and $\theta$ is the angle between the link and the horizontal
line (see figure~\ref{fig:SingleLinkLinearTrack}).

The Euler Lagrange equation of motion reads:
 \begin{equation}
  \label{eq:betterBeadEoM}
 {d \over d\theta}({\dot{s}\over\sqrt{\dot{s}^2}} + {\dot{s} + cos\theta \over
    \sqrt{\dot{s}^2 + 1 + 2\dot{s}cos\theta}}) = 0
\end{equation}

We consider a transformation which is not (necessarily) a minimum:
\begin{equation}
  \label{eq:1linkhyperexsol}
  s = a \cos \theta - \sin \theta + b
\end{equation}
with $a$ and $b$ parameters to be determined. 

Such a transformation in fact forces the two ends to travel on a
straight line (right from the beginning) , but the $A$ side may in
fact retreat and then move forward. We call such a
transformation a ``hyperextended transformation''.  A sample
transformation of this kind is shown in figure
\ref{fig:SingleLinkLinearTrack}. The parameters $a$ and $b$ in
(\ref{eq:1linkhyperexsol}) can be tuned to meet the boundary
conditions (see below).

In fact it is seen that point $A$ on the link retreats backwards until
it reaches some critical angle, which is when link $\ovl{AB}$ makes an
angle ${\pi \over 2}$ with the straight line $\ovl{BB'}$ that point
$B$ travels on. Subsequently $A$ then moves forward towards $A'$.

Assume that $\theta$ runs from $\theta_1$ to $\th_2$, where
$0<\theta_2<\pi/2$. For 
simplicity assume that both these angles are between $0$ and ${\pi
  \over 2}$.

The boundary conditions dictate that:
\begin{eqnarray}
  s(\theta_1) &=& 0\\
  s(\theta_2) &=& l
\end{eqnarray}
where $l$ is the distance between $A$ and $A^{\prime}$.

$a$ and $b$ can be explicitly solved to give:
\begin{eqnarray}
  \label{eq:solveforA}
  a &=&{{-\sin \theta_2 + \sin \theta_1-l}\over{\cos \theta_1-
      \cos \theta_2}}\\
  \label{eq:solveforB}
  b &=&-{{\cos \theta_1\,\left(-\sin \theta_2-l\right)+\sin \theta_1\,\cos \theta_2}\over
    {\cos \theta_1-\cos \theta_2}}
\end{eqnarray}

For our purposes we only need to note
that the critical angle occurs when $\dot{s} \equiv {d s \over d \theta}$ becomes zero,
that is when $A$ stops going backward and starts moving forward:
\begin{equation}
  \label{eq:critanglederive}
  \dot s = -a \sin \theta - \cos \theta = 0
\end{equation}
where $a$ is given in \ref{eq:solveforA}.

We can now ask what should $\theta_1$ be so that there is
no need for the link to go backward, i.e. it moves forward
from the beginning and the transformation is monotonic.
Equations~(\ref{eq:critanglederive}) and~(\ref{eq:solveforA}) give:
\begin{equation}
  \label{eq:CriticalAngle}
  \cos \theta + {{-\sin \theta_2 + \sin \theta-l}\over{\cos \theta-
      \cos \theta_2}} \sin \theta = 0 
\end{equation}

\begin{figure}[h]
  \centering
  \includegraphics[width=4cm]{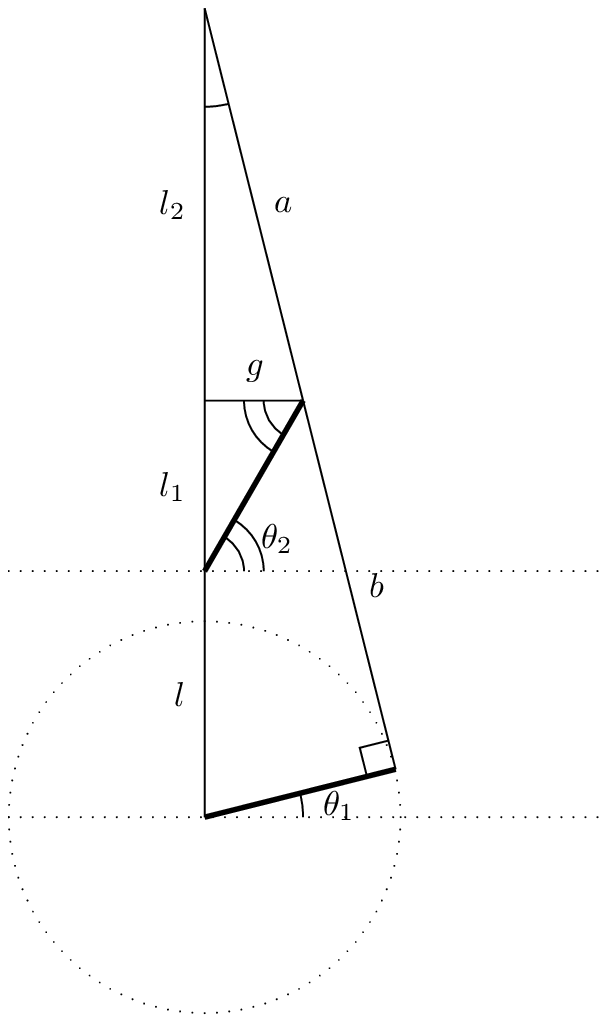}
  \caption{Geometric proof for critical angle condition}
  \label{fig:geoLinkGeneral}
\end{figure}

For pedagogical reasons we prove condition~(\ref{eq:CriticalAngle})
using analytic geometry as well. Looking at
figure~\ref{fig:geoLinkGeneral} we  
have the following:
\begin{eqnarray}
  \label{eq:genGeoProofp1}
  g^2 + l_1^2 &=& 1 \\
  g^2 + l_2^2 &=& a^2 \\
  {g \over a} &=& {1 \over l + l_1 + l_2}
\end{eqnarray}
We can solve $g = \sqrt{1 - l_1^2}$ and $a = \sqrt{1 - l_1^2 + l_2^2}$ from
the first two equations and substitute in the third equation to give:
\begin{equation}
  \label{eq:genGeoChF}
  l = {\sqrt{1-l_1^2 +l_2^2} \over \sqrt{1 - l_1^2}} - l_1 - l_2
\end{equation}
On the other hand based on our results for $g$ and $a$ we have:
\begin{eqnarray}
  \label{eq:genGeoProofp2}
  \sin \theta_1 &=& {\sqrt{1 - l_1^2} \over \sqrt{1 - l_1^2 + l_2^2}}\\
  \cos \theta_1 &=& {l_2 \over \sqrt{1 - l_1^2 + l_2^2}}\\
  \sin \theta_2 &=& l_1\\
  \cos \theta_2 &=& \sqrt{1 - l_1^2}
\label{eq:genGeoProofp2end}
\end{eqnarray}
Substitution of
eqns~(\ref{eq:genGeoProofp2}-\ref{eq:genGeoProofp2end}) in
equation~(\ref{eq:CriticalAngle}) gives equation~(\ref{eq:genGeoChF})
after some simplification.

For the particular case that we have discussed, the proposed
transformation is in fact a minimal solution if $\theta_1$ is greater
than the critical angle, because in that case a simple slide would be
possible. If $\theta_1$ is less than the critical angle a locally
minimum solution as we know is pure rotation to the critical angle and
then straight line slide. Pure rotation has a nice geometric
interpretation in our parametrization. it corresponds to the null
solution $s=0$. Since at the critical angle $\dot{s} = 0$ we see that
$s=0$ will be smoothly patched with $ s = a \cos \theta - sin \theta +
b$, as mandated by the corner conditions in
equation~(\ref{eq:cornercond1}).  

\begin{figure}[h]
  \centering
  \includegraphics[width=6cm]{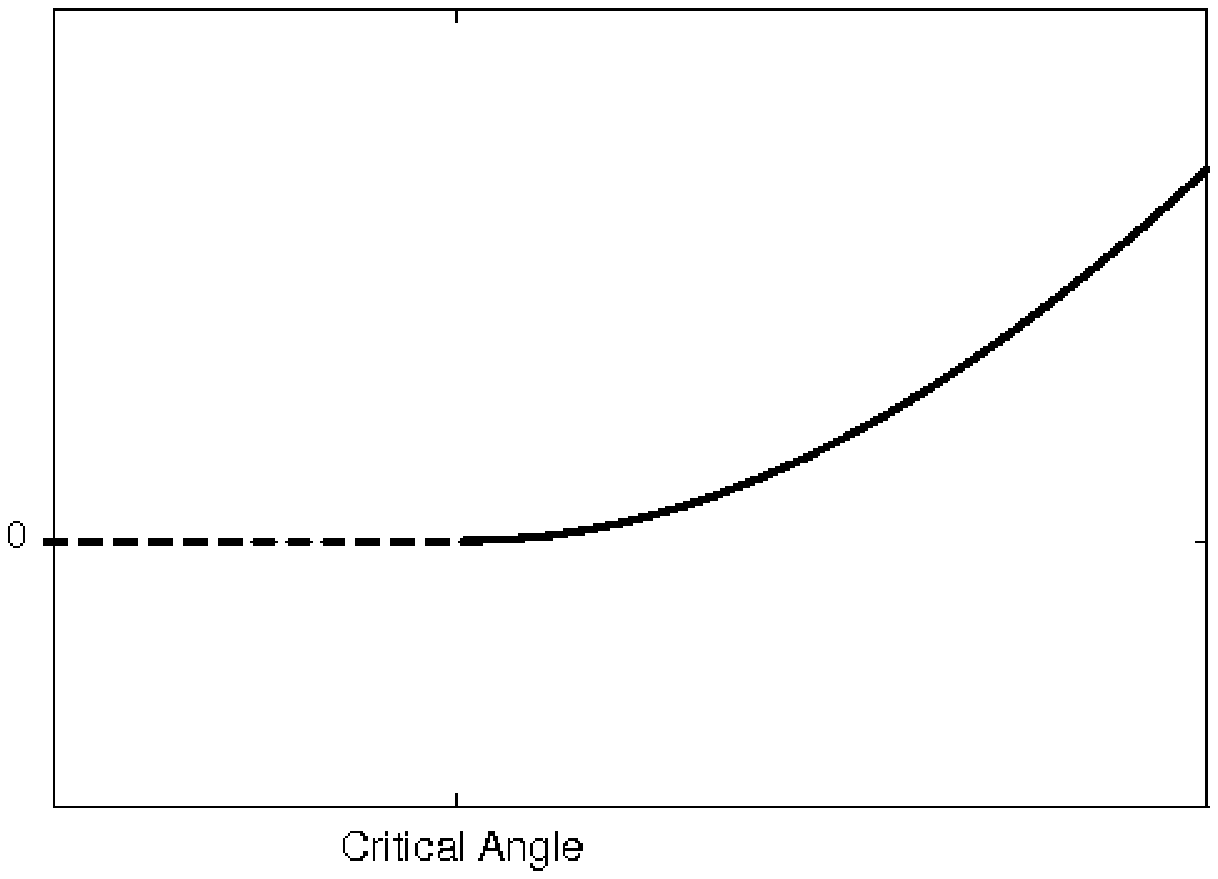}
  \caption{A minimal transformation in $s(\theta)$ parametrization.
    The horizontal segment corresponds to pure rotation and the curved
    section corresponds to slide on straight paths. Here the corner
    conditions demand that the derivative $\dot{s}$ be continuous at
    the critical angle. }
  \label{softtheta}
\end{figure}

\section{Minimal transformations in 2 dimensions}
\label{app:min2d}

It was seen in section~\ref{sec:2linkconvexity} that for the case of
two links when one is confined
to moving in a plane, satisfying the constant link length
constraints and corner conditions do not seem to lead to solutions
which are extremal. However given the additional constraint that the
links must lie in a plane, there must be one or a set of minimal
transformations.  We need to look at other forms of
transformations, namely compound straight line transformations. We will
elaborate on the idea starting with single links.

\begin{figure}[h]
  \centering
  \includegraphics[width=4.5cm]{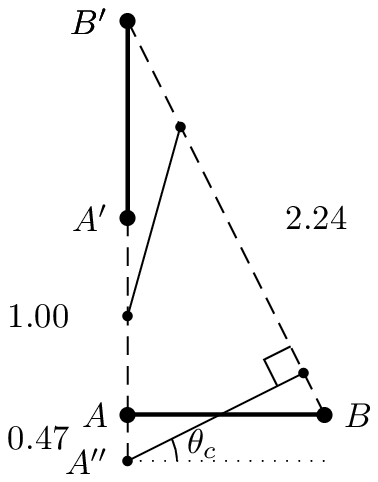}
  \includegraphics[width=4.5cm]{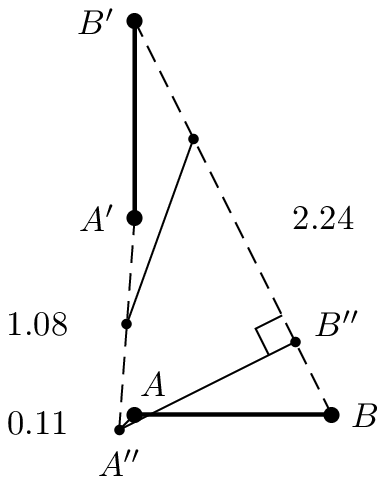}
  \myCaption{
The previous hyper extended solution is shown along with a
    more general compound straight-line transformation, where
    $\ora{AA''}$ travels in some general direction. Length of
    each line segment is written beside it. For the hyper extended
    solution the value of $AA''$ is multiplied by two
    because the path is travelled twice.}
  \label{fig:1linkHyperExDegen}
\end{figure}

The hyper extended solution that was discussed previously
in~\ref{app:critangles}  
can be considered as a very special example of compound straight line
transformation. These are transformations that are made strictly from
straight line paths with no pure rotation. A more general
transformation is shown figure \ref{fig:1linkHyperExDegen} beside the
old transformation.

Note that the corners do not technically violate the corner conditions
because the speed of ``A'' bead is zero at the corner point in any
parametrization that can simultaneously describe $A$ motion and $B$
motion: Since at the corner point, the link makes an angle of 90
degrees with the path that B travels, the speed of B at the critical
angle in infinitely larger than the speed of A. In fact one sees that
we have an instantaneous pure rotation about $A$-bead, when it is at
the corner point.  $\hat{v}_a$ is not clearly defined at the corners,
and everywhere else (when the speed of the bead(s) is not zero), the
two beads are travelling on a straight line. The two solutions
depicted in the figure come from two different parametrization of the
most general form of the action and result in different distances.
But each of them is a local minimum once the direction of $\ora{AA''}$ is
picked, and these local minima have different values for the distance.

We can then ask about the best position to put the corner point, to minimize
the distance travelled in the compound straight line transformation,
with respect to other compound straight line transformations.
We assume the corner occurs on one side and we take it to be the ``A''
side. 

Note that at the corner, the link makes a $90^\circ$ angle with the
B-bead path $\ovl{BB'}$, meaning that the distance from the corner
point to B path 
is always the length of the link, i.e. unity here. 
Also note that the total
distance that the ``A''-bead travels is the distance from the initial
point $A$ to the corner point $A''$, plus the distance from
$A''$ to the final position $A^{\prime}$. 

\begin{figure}[h]
  \centering
  \includegraphics[width=5cm]{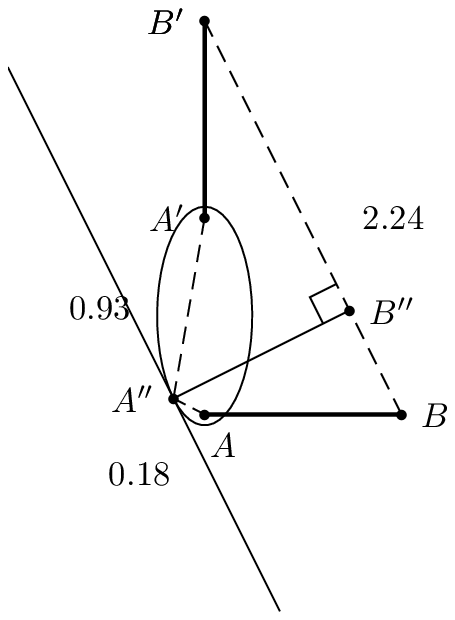}
  \caption{Optimal Compound Straight Line Transformation}
  \label{fig:optimalhyperex}
\end{figure}

The locus of points with equal sum of distances from two points $A$
and $A'$ defines an ellipse with foci at $A$ and $A'$. Moreover the
length of the major axis of the ellipse equals the sum of the
distances from the foci.  Thus the smaller the major axis of the
ellipse with foci $A$ and $A^{\prime}$, the smaller the total distance
travelled by the ``A''-bead.  Moreover $A''$ should sit on a line
parallel to B-path at a distance of $1$ from the B-path line
$\ovl{BB'}$.  So in seeking the shortest distance travelled the $A$
end of the link, we seek the point $A^{\prime\prime}$ such that it
lies on an ellipse with foci $A$ and $A^{\prime}$, the ellipse shares
at least one point with a line parallel to $\ovl{BB'}$ and distance
$1$ away from it, and lastly that the ellipse has the smallest possible
major axis (see figure~\ref{fig:optimalhyperex}).  So the ellipse
giving the minimal distance is tangent to the parallel line, and
$A^{\prime\prime}$ is the tangent point.  This is illustrated in
figure~\ref{fig:optimalhyperex}.

\begin{figure}[h]
  \centering
  \includegraphics[width=6cm]{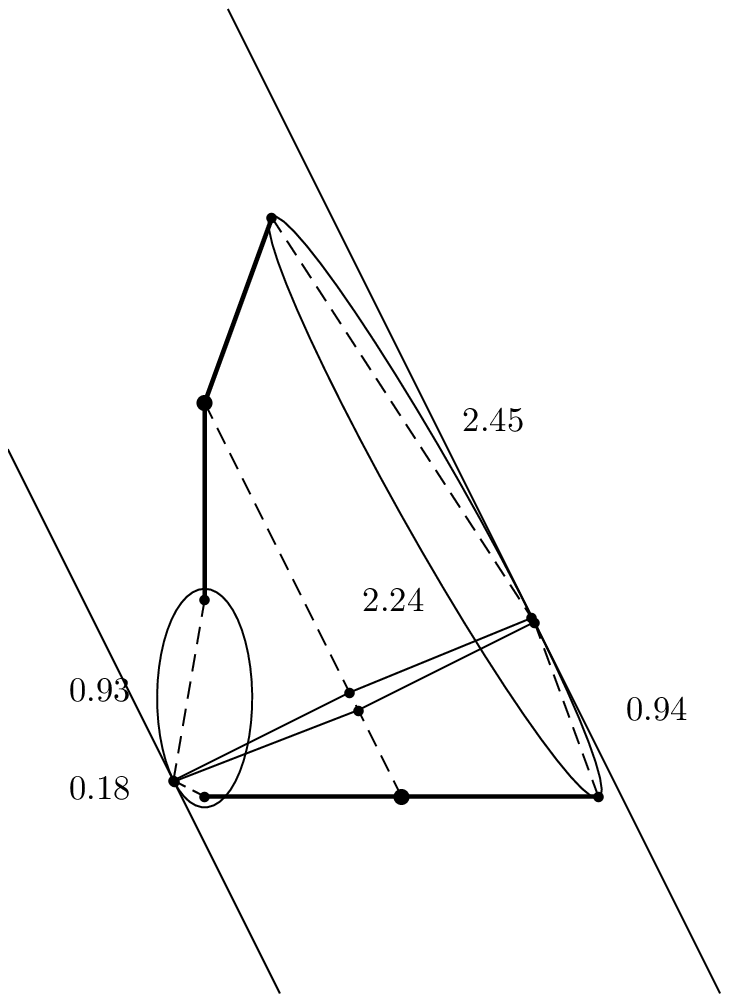}
  \caption{An optimal compound Straight line solution for 2 link. For
    this particular class of solutions, the problem is divided into to disjoint
    problems (one for each link) and solved separately.}
  \label{fig:2linkopthyperex}
\end{figure}

\begin{figure}
  \centering
  \includegraphics[width=6cm]{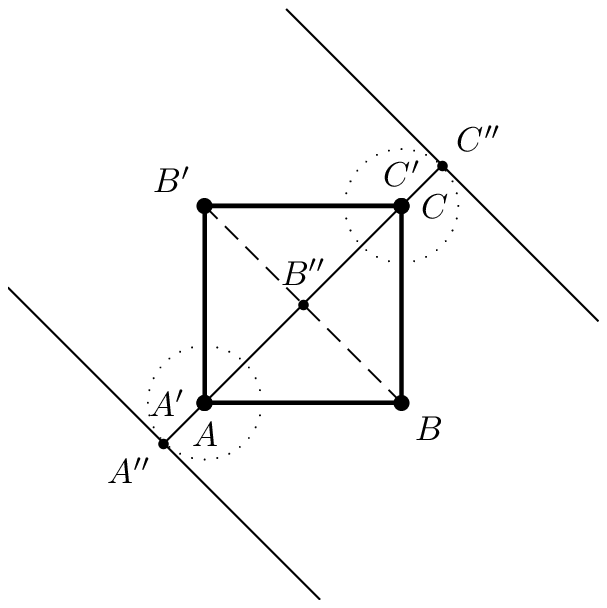}
  \caption{Minimal transformation restricted to 2 dimensions, for 2 links of
    opposite convexity 
    which form opposite sides of a square.}
  \label{squarefig}
\end{figure}

This solution can be straightforwardly extended to 2 links, as
depicted in figure 
\ref{fig:2linkopthyperex}.
Consider then the example in figure~\ref{limlink}a, where the links
are no longer allowed to move out of the plane (see
figure~\ref{squarefig}). Here  
$\bfr_A = \bfr_{A^{\prime}}$ and $\bfr_C = \bfr_{C^{\prime}}$ and the
above ellipses turns into a circles centered at $A$ and $C$. 
The circles have radii $1-1/\sqrt{2}$, so that the perpendicular
distance from line 
$\ovl{BB'}$ to the farthest point on the circle is $1$ and a fully
extended intermediate state is allowed.


\newpage
\hspace{5cm} \line (1,0){200} \\
{\bf References} \\

\bibliographystyle{my_pnas}

.

\end{document}